%% file: 00_main.tex
\documentclass[journal, xcolor=table]{IEEEtran}
\makeatletter
\def\endthebibliography{%
	\def\@noitemerr{\@latex@warning{Empty `thebibliography' environment}}%
	\endlist
}
\makeatother
\usepackage{amsmath,amssymb,amsthm}
\usepackage{cite}
\usepackage[hyphens]{url}
\usepackage[hidelinks, bookmarks=false, final=false]{hyperref}

\usepackage[capitalize ,nameinlink]{cleveref} 
\renewcommand{\autoref}[1]{\Cref{#1}} %% Just used for the transistion from hyperref to cleaverref package. Use \Cref when refering to ex. Figure 4.1

\Crefname{equation}{Eq.}{Eqs.}
\Crefname{figure}{Fig.}{Figs.}
\Crefname{tabular}{Tab.}{Tabs.}

   \usepackage[pdftex]{graphicx}
\usepackage{amsmath}
\usepackage[utf8]{inputenc}					% Input-indkodning af tegnsaet (UTF8)

\usepackage{algorithmic}
\usepackage{multirow}
\usepackage{multicol}

\usepackage{subcaption}

\usepackage{float}

\usepackage[binary-units=true]{siunitx}

\usepackage[table]{xcolor}
\usepackage{tabularx}
\usepackage{makecell}

\usepackage{tikz}
\usetikzlibrary[patterns]
\usetikzlibrary{arrows}
\usepackage[siunitx]{circuitikz}
\usetikzlibrary{matrix,decorations.pathmorphing,shapes}

\usepackage[xindy,acronym,shortcuts,nonumberlist,style=long,savewrites=true]{glossaries} 

\usepackage{mathtools}	
\usepackage{bm}	

\makenoidxglossaries
\glsenableentrycount % enable \cgls, \cglspl, \cGls, \cGlspl (Used for detecting if a gls only is used once)

\usepackage{todonotes}

\newcommand{\glsl}[1]{% \glsl, gls long entry
	\glsentrylong{#1}\unskip
}% 
\newcommand{\glss}[1]{% \glsl, gls long entry
	\glsentryshort{#1}\unskip
}% 
\DeclarePairedDelimiter\ceil{\lceil}{\rceil}
\DeclarePairedDelimiter\floor{\lfloor}{\rfloor}

\makeatletter
\providecommand\add@text{}
\newcommand\eqUnit[1]{%
  \gdef\add@text{#1\gdef\add@text{}}}% 
\renewcommand\tagform@[1]{%
  \maketag@@@{\llap{\add@text\quad}(\ignorespaces#1\unskip\@@italiccorr)}%
}
\makeatother

\input{_glossary}
\begin{document}

\bstctlcite{IEEEexample:BSTcontrol} % Avoid repeated authors to make into ----- see IEEEexample.bib for the control code.

% \title{An Empirical Estimation of the Battery Lifetime of LTE-M and NB-IoT Devices}
% \title{An Empirical Investigation of the Battery Lifetime of LTE-M and NB-IoT Devices}

% \title{A Comprehensive, Measurement-based Analysis of the Battery Lifetime of LTE-M and NB-IoT Devices}
\title{Modelling and Experimental Validation for Battery Lifetime Estimation in NB-IoT and LTE-M}




%%%%%% AUTHORS %%%%
%% Author two block
% \author{\IEEEauthorblockN{André Sørensen, Maxime Jerome Remy, Nicolaj Kjettrup, Rasmi Vlad Mahmoud}
% 	\IEEEauthorblockA{Department of Electronic Systems, Aalborg University, Aalborg, Denmark\\
% 		\{asare14, mremy16, nkjett14, rmahmo17\}@student.aau.dk}
% 		\\
% 		\IEEEauthorblockN{Jens Myrup Pedersen}
% 	    \IEEEauthorblockA{Department of Electronic Systems, Aalborg University, Aalborg, Denmark\\
% 		jens@es.aau.dk}}
% IEE membership: ~\IEEEmembership{Member,~IEEE,}
%% Author one block 
\author{\IEEEauthorblockN{André Sørensen\IEEEauthorrefmark{1}, Hua Wang\IEEEauthorrefmark{1}\IEEEauthorrefmark{3}, Maxime Jérôme Remy\IEEEauthorrefmark{1}, Nicolaj Kjettrup\IEEEauthorrefmark{1}, Ren\'{e}~Brandborg~S{\o}rensen\IEEEauthorrefmark{2}, \\ Jimmy Jessen Nielsen\IEEEauthorrefmark{2}, Petar Popovski\IEEEauthorrefmark{2}, Germán Corrales Madueño\IEEEauthorrefmark{1}\IEEEauthorrefmark{2}}
	\IEEEauthorblockA{\\\IEEEauthorrefmark{1}Keysight Technologies Denmark Aps, Aalborg, Denmark\\ 
 \IEEEauthorrefmark{1}\{andre.soerensen,maxime.remy,nicolaj.kjettrup,german.madueno\}@keysight.com\\ \IEEEauthorrefmark{2}Department of Electronic Systems, Aalborg University, Aalborg, Denmark\\
 \IEEEauthorrefmark{2}\{jjn, petarp\}@es.aau.dk
 \\\IEEEauthorrefmark{3}Zhejiang Lab, Hangzhou, China\\
 \IEEEauthorrefmark{3}hua.wang@zhejianglab.com
		}}
% \and
% \IEEEauthorblockN{Germán Corrales Madueño}
% \IEEEauthorblockA{Keysight Technologies Denmark Aps \\ Aalborg, Denmark\\
% german.madueno@keysight.com\\}}
%%%% AUTHOR END %%%%%

%


\maketitle

\begin{abstract}
\cgls{iot} is one of the main features in 5G. \cgls{lpwan} has attracted enormous research interests to enable large scale deployment of \cgls{iot}, with the design objectives of low cost, wide coverage area, as well as low power consumption. In particular, long battery lifetime is essential since many of the \cgls{iot} devices will be deployed in hard-to-access locations. Prediction of the battery lifetime depends on the accurate modelling of energy consumption. This paper presents a comprehensive power consumption model for battery lifetime estimation, which is based on \cgls{ue} states and procedures, for two cellular \cgls{iot} technologies: \cgls{nbiot} and \cgls{ltem}. A measurement testbed has been setup and the proposed model has been tested and validated via extensive measurements under various traffic patterns and network scenarios, achieving the modelling inaccuracy within 5\%. The measurement results show that the battery lifetime of an \cgls{iot} device can reach up to 10 years as required by 3GPP, with proper configuration of the traffic profile, the coverage scenario, as well as the network configuration parameters.
\end{abstract}


% \fi
\IEEEpeerreviewmaketitle

\input{01_Introduction_new}

\input{02_PowerModel.tex}

\input{03_Scenario.tex}

\input{04_Results.tex}

\input{05_Conclusion.tex}

\bibliographystyle{Setup/IEEEtran}
\bibliography{Bib/IEEEexampleNew}

% \begin{appendices}
% \input{AppendixLTEM.tex}
% \end{appendices}

\end{document}

%% file: _glossary.tex
\newacronym{n2}{N211}{u-blox SARA-N211-02B}
\newacronym{r4}{R412M}{u-blox SARA-R412M-02B}
\newacronym{bc95}{BC95}{Quectel BC95-B20}

\newacronym{emmd}{\glss{emm}\_DEREGISTERED}{\glss{emm}\_DEREGISTERED}
\newacronym{emmr}{\glss{emm}\_REGISTERED}{\glss{emm}\_REGISTERED}
\newacronym{ecmi}{\glss{ecm}\_IDLE}{\glss{ecm}\_IDLE}
\newacronym{ecmc}{\glss{ecm}\_CONNECTED}{\glss{ecm}\_CONNECTED}
\newacronym{rrci}{\glss{rrc}\_IDLE}{\glss{rrc}\_IDLE}
\newacronym{rrcc}{\glss{rrc}\_CONNECTED}{\glss{rrc}\_CONNECTED}

\newacronym{s1ap}{S1-\glss{ap}}{S1 \glsl{ap}}
\newacronym{lorawan}{LoRaWAN}{Long Range Wide Area Network}
\newacronym{sigfox}{Sigfox}{Sigfox}
\newacronym{wifi}{Wi-Fi}{Wi-Fi}

\newacronym{vs}{VS.}{Versus}

\newacronym{cagr}{CAGR}{compound annual growth rate}
\newacronym{ntn}{NTN}{non terrestrial networks}
\newacronym{enb}{eNB}{E-UTRAN Node B}
\newacronym{fenb}{FeNB}{Femto eNB}
\newacronym{mqtt}{MQTT}{Message Queuing Telemetry Transport}
\newacronym{tap}{TAP}{Test Automation Platform}
\newacronym{na}{N/A}{Not Applicable}
\newacronym{aau}{AAU}{Aalborg University}
\newacronym{3g}{3G}{$3^{rd}$ Generation}
\newacronym{3gpp}{3GPP}{3rd Generation Partnership Project}
\newacronym{lte}{LTE}{Long Term Evolution}
\newacronym{ltem}{LTE-M}{\glsentrylong{lte} for Machines}
\newacronym{iot}{IoT}{Internet of Things}
\newacronym{ciot}{CIoT}{Cellular \glsentrylong{iot}}
\newacronym{nbiot}{NB-IoT}{Narrowband \glsentrylong{iot}}
\newacronym{gsm}{GSM}{Global System for Mobile Communications}
\newacronym{edge}{EDGE}{Enhanced Data Rates for GSM Evolution}
\newacronym{dl}{DL}{Downlink}
\newacronym{ul}{UL}{Uplink}
\newacronym{ofdma}{OFDMA}{Orthogonal Frequency-Division Multiple Access}
\newacronym{ofdm}{OFDM}{Orthogonal Frequency-Division Multiplexing}
\newacronym{tdma}{TDMA}{Time Division Multiple Access}
\newacronym{scfdma}{SC-FDMA}{Single-Carrier Frequency-Division Multiple Access}
\newacronym{pa}{PA}{Power Amplfier}
\newacronym{fdma}{FDMA}{Frequency-Division Multiple Access}
\newacronym{mac}{MAC}{Medium Access Control}
\newacronym{phy}{PHY}{Physical}
\newacronym{qos}{QoS}{Quality of Service}
\newacronym{qci}{QCI}{\glss{qos} Class Identifier}
\newacronym{as}{AS}{Access Stratrum}
\newacronym{nas}{NAS}{Non-Access Stratrum}
\newacronym{pcc}{PCC}{Policy and Charging Control}
\newacronym{af}{AF}{Application Function}
\newacronym{emm}{EMM}{\glss{eps} Mobility Management}
\newacronym{esm}{ESM}{\glss{eps} Session Management}
\newacronym{ecm}{ECM}{\glss{eps} Connection Management}
\newacronym{ce}{CE}{Coverage Enhancement}
\newacronym{snr}{SNR}{Signal to Noise Ratio}
\newacronym{nb}{NB}{Narrowband}
\newacronym{gprs}{GPRS}{General Packet Radio Service}
\newacronym{gtp}{GTP}{\glss{gprs} Tunneling Protocol}
\newacronym{gbr}{GBR}{Guaranteed Bit Rate}
\newacronym{rat}{RAT}{Radio Access Technology}
\newacronym{psu}{PSU}{Power Supply Unit}
\newacronym{tx}{TX}{Transmission}
\newacronym{rx}{RX}{Reception}
\newacronym{fdd}{FDD}{Frequency-Division Duplex}
\newacronym{tdd}{TDD}{Time-Division Duplex}
\newacronym{fdfdd}{FD-FDD}{Full-Duplex \glsentryshort{fdd}}
\newacronym{hdfdd}{HD-FDD}{Half-Duplex \glsentryshort{fdd}}
\newacronym{harq}{HARQ}{Hybrid Automatic Repeat Request}
\newacronym{fec}{FEC}{Forward Error Correction}
\newacronym{arq}{ARQ}{Automated Repeat reQuest}
\newacronym{crc}{CRC}{Cyclic Redundancy Check}
\newacronym{ack}{ACK}{Acknowledgement}
\newacronym{nack}{NACK}{Negative Acknowledgement}
\newacronym{tau}{TAU}{Tracking Area Update}
\newacronym{rau}{RAU}{Routing Area Update}
\newacronym{dci}{DCI}{Downlink Control Information}
\newacronym{uci}{UCI}{Uplink Control Information}
\newacronym{imsi}{IMSI}{International Mobile Subscriber Identity}
\newacronym{pcid}{PCID}{Physical Cell Identity}
\newacronym{mib}{MIB}{Master Information Block}
\newacronym{sib}{SIB}{System Information Block}
\newacronym{ptw}{PTW}{Paging Time Window}
\newacronym{csi}{CSI}{Channel Status Information}
\newacronym{sr}{SR}{Scheduling Request}
\newacronym{ra}{RA}{Random Access}
\newacronym{bler}{BLER}{Block Error Rate}
\newacronym{tm}{TM}{Transmission Mode}
\newacronym{amcs}{AMCS}{Adaptive Modulation and Coding Scheme}
\newacronym{mcs}{MCS}{Modulation and Coding Scheme}
\newacronym{tb}{TB}{Transport Block}
\newacronym{tbs}{TBS}{Transport Block Size}
\newacronym{mtc}{MTC}{Machine-Type Communication}
\newacronym{ciotepsopt}{\glss{ciot}~\glss{eps}~Optimization}{Cellular~\glss{iot}~\glss{eps}~Optimization}
\newacronym{pdn}{PDN}{Packet Data Network}
\newacronym{voip}{VoIP}{Voice over IP}
\newacronym{utran}{UTRAN}{Universal Terrestrial Radio Access Network} 
\newacronym{geran}{GERAN}{\glss{gsm} \glss{edge} Radio Access Network} 
\newacronym{mt}{MT}{Mobile Terminated}
\newacronym{mo}{MO}{Mobile Originated}
\newacronym{ap}{AP}{Application Protocol}
\newacronym{pdu}{PDU}{Protocol Data Unit}
\newacronym{rai}{RAI}{Release Assistance Information}
\newacronym{msg}{MSG}{Message}
\newacronym{ru}{RU}{Resource Unit}
\newacronym{rf}{RF}{Radio frequency}
\newacronym{am}{AM}{Acknowledged Mode}
\newacronym{rmse}{RMSE}{Root Mean Square Error}
\newacronym{tpc}{TPC}{Target Power Control}
\newacronym{pp}{pp}{percentage point}
\newacronym{qpsk}{QPSK}{Quadrature Phase Shift Keying}
\newacronym{bpsk}{BPSK}{Binary Phase Shift Keying}
\newacronym{16qam}{16-QAM}{16 bits Quadrature Amplitude Modulation}
\newacronym{ms}{MS}{Modulation Scheme}
\newacronym{mbb}{MBB}{Mobile Broadband}
\newacronym{nfc}{NFC}{Near-field communication}
\newacronym{rfid}{RFID}{Radio-frequency identification}
\newacronym{lpwan}{LPWAN}{Low-power wide-area networking}
\newacronym{dut}{DUT}{device-under-test}

\newacronym{tp}{TP}{Traffic Profile}

\newacronym{SubSpace}{$\Delta f$}{Sub-carrier Spacing}

\newacronym{tcp}{TCP}{Transmission Control Protocol}
\newacronym{ip}{IP}{Internet Protocol}
\newacronym{mcl}{MCL}{Maximum Coupling Loss}
\newacronym{cl}{CL}{Coupling Loss}
\newacronym{bw}{BW}{Bandwidth}

\newacronym{hsfn}{H-SFN}{Hyper System Frame Number}
\newacronym{sfn}{SFN}{System Frame Number}
\newacronym{sn}{SN}{Subframe Number}
\newacronym{uss}{USS}{\glss{ue} specific Search Space}
\newacronym{css}{CSS}{Common Search Space}
\newacronym{sf}{SF}{Subframe}
\newacronym{re}{RE}{Ressource Element}

\newacronym{ue}{UE}{User Equipment}

\newacronym{drx}{DRX}{Discontinuous Reception}
\newacronym{edrx}{eDRX}{Extended \glsl{drx}}
\newacronym{idrx}{iDRX}{Idle Mode \glsl{drx}}
\newacronym{cdrx}{cDRX}{Connected Mode \glsl{drx}}
\newacronym{ecdrx}{c-eDRX}{Connected Mode \glsl{edrx}}
\newacronym{psm}{PSM}{Power Saving Mode}
\newacronym{eps}{EPS}{Evolved Packet System}
\newacronym{eps-aka}{EPS-AKA}{\glsentrylong{eps} Authentication and Key Agreement}
\newacronym{guti}{GUTI}{Globally Unique Temporary Identity}
\newacronym{ta}{TA}{Tracking Area}
\newacronym{cqi}{CQI}{Channel Quality Indicator}
\newacronym{mimo}{MIMO}{Multiple Inputs Multiple Outputs}
\newacronym{ri}{RI}{Rank Indicator}
\newacronym{pmi}{PMI}{Precoding Matrix Indicator}
\newacronym{prb}{PRB}{Physical Resource Block}
\newacronym{rb}{RB}{Resource Block}

\newacronym{pss}{PSS}{Primary Synchronization Signal}
\newacronym{sss}{SSS}{Secondary Synchronization Signal}
\newacronym{pbch}{PBCH}{Physical Broadcast Channel}
\newacronym{pcfich}{PCFICH}{Physical Control Format Indicator Channel}
\newacronym{pdcch}{PDCCH}{Physical Downlink Control Channel}
\newacronym{pdsch}{PDSCH}{Physical Downlink Shared Channel}
\newacronym{phich}{PHICH}{Physical Hybrid ARQ Indicator Channel}

\newacronym{pucch}{PUCCH}{Physical Uplink Control Channel}
\newacronym{pusch}{PUSCH}{Physical Uplink Shared Channel}
\newacronym{prach}{PRACH}{Physical Random Access Channel}

\newacronym{mpdcch}{MPDCCH}{MTC Physical Downlink Control Channel}
\newacronym{ereg}{EREG}{Enhanced Ressource Element Group}
\newacronym{ecce}{ECCE}{Enhanced Channel Control Elements}

\newacronym{npss}{NPSS}{Narrowband Primary Synchronization Signal}
\newacronym{nsss}{NSSS}{Narrowband Secondary Synchronization Signal}
\newacronym{npbch}{NPBCH}{Narrowband Physical Broadcast Channel}
\newacronym{npdcch}{NPDCCH}{Narrowband Physical Downlink Control Channel}
\newacronym{npdsch}{NPDSCH}{Narrowband Physical Downlink Shared Channel}
\newacronym{nprach}{NPRACH}{Narrowband Physical Random Access Channel}
\newacronym{npusch}{NPUSCH}{Narrowband Physical Uplink Shared Channel}

\newacronym{ncce}{NCCE}{Narrowband Control Channel Element}

\newacronym{e-utran}{E-UTRAN}{Evolved Universal Terrestrial Radio Access Network}
\newacronym{enodeb}{eNodeB}{\glss{e-utran} Node B}
\newacronym{epc}{EPC}{Evolved Packet Core}
\newacronym{cp}{CP}{Control Plane}
\newacronym{up}{UP}{User Plane}
\newacronym{rrm}{RRM}{Radio Ressource Management}
\newacronym{mme}{MME}{Mobility Management Entity}
\newacronym{sgw}{S-GW}{Serving Gateway}
\newacronym{pgw}{P-GW}{Packet data network Gateway}
\newacronym{hss}{HSS}{Home Subscription Server}
\newacronym{pcrf}{PCRF}{Policy and Charging Resource Function}
\newacronym{rlc}{RLC}{Radio Link Control}
\newacronym{pdcp}{PDCP}{Packet Data Convergence Protocol Layer}
\newacronym{rrc}{RRC}{Radio Resource Control}
\newacronym{pmt}{PMT}{Power Measurement Tool}
\newacronym{gui}{GUI}{Graphical User Interface}

\newacronym{rap}{RAP}{Random Access Preamble}
\newacronym{rar}{RAR}{Random Access Response}
\newacronym{rrcReq}{\cgls{rrc} Connection Request}{\cgls{rrc} Connection Request}
\newacronym{rrcSet}{\cgls{rrc} Connection Setup}{\cgls{rrc} Connection Setup}
\newacronym{rrcComp}{\cgls{rrc} Connection Complete}{\cgls{rrc} Connection Setup Complete}
\newacronym{rrcResComp}{\cgls{rrc} Resume Complete}{\cgls{rrc} Connection Resume Complete}
\newacronym{rrcResReq}{\cgls{rrc} Resume Request}{\cgls{rrc} Resume Request}
\newacronym{rrcRes}{\cgls{rrc} Connection Resume}{\cgls{rrc} Connection Resume}

%% file: 01_Introduction_new.tex
\section{Introduction}\label{sec:intro}
The number of \cgls{iot} connections has increased exponentially worldwide since 2015, reaching over 1 billion in 2020, and is predicted to keep growing to  roughly 5 billion by 2025 \cite{ericssonMobilityReport}.
Early \cgls{iot} deployments have been used for smart metering, asset monitoring and smart logistics, industrial automation, traffic monitoring, fleet tracking, smart home appliances, etc. The \cgls{iot} applications are designed to support massive deployment, e.g., more than 100,000 per cell. Replacing the batteries for such a large number of devices would not only be costly and cumbersome, but also sometimes impractical, since devices may be deployed in locations where it is hard for human to access, for example in basement where the water meters are placed. It often ends up that the battery lifetime determines the lifetime of the \cgls{iot} device. Therefore, massive connectivity, low cost, wide coverage, and low power consumption are the key requirements for the \cgls{iot} devices.

To address the requirements imposed by \cgls{iot} applications, \gls{3gpp} has introduced two standards for \cgls{ciot}: \cgls{ltem} and \cgls{nbiot}, both of which are developed based on 4G \cgls{lte}, but are targeted for \gls{lpwan} solutions. The main advantages of \gls{3gpp} standardized technologies as compared to other proprietary solutions are the reuse of existing infrastructure and the use of a licensed spectrum, allowing for more stable and predictable performance. 

Energy efficiency is certainly a major concern for typical \cgls{iot} deployment, especially for battery-powered \cgls{iot} devices, which are normally affected by a vast number of network configurations as well as implementation parameters. \cgls{3gpp} requires a battery lifetime of more than 10 years with a battery capacity of 5 Wh for an \cgls{iot} device operating with a predefined traffic profile in certain deployment scenarios \cite{TR45820}. Therefore, accurate modelling of the energy consumption for an \cgls{iot} device is of great importance, as the application developers as well as the network operators need to predict and plan which \cgls{iot} technology or device should be chosen and how the network parameters should be configured based on the performance requirements and the expected battery lifetime for a given use case. The model should be general enough so that it can be applied to different devices running on same or different \cgls{iot} technologies. The model should also be flexible enough with different configurations of network and application parameters, for example various traffic profile and network scenarios, and should be able to analyze the impact of different configurations on the device performance.

Early studies on the modelling of power consumption and the empirical performance characterization with respect to \cgls{nbiot} are discussed in details in Section~\ref{sec:relatedwork}. A preliminary study on the power consumption model specifically for \cgls{nbiot} was presented in our earlier work \cite{iotbook}. Comparatively, this paper improves the power consumption model proposed in \cite{iotbook} and adapts the analysis to also include \cgls{ltem}. Additionally, the proposed model has been validated via empirical power consumption measurements and the impact of network and application configuration parameters on the device’s battery lifetime has been analyzed. Specifically, the main contributions of this paper are summarized below:
\begin{enumerate}
  \item We extend the power consumption model to include \cgls{ltem}. The proposed model can be generalized to estimate the energy consumption and the battery lifetime of any devices running either \cgls{nbiot} or \cgls{ltem}, with flexible configurations of traffic profile, coverage scenario, as well as network parameters.
  \item The accuracy of the model has been improved based on PHY-level measurements and detailed modelling of each \cgls{ue} state and procedure. By composing different components of \cgls{ue} states and procedures, the power consumption of any \cgls{ue} behaviour can be modelled. Extensive measurements have been performed using two commercial \cgls{iot} devices with the latest \cgls{nbiot}/\cgls{ltem} features to validate the proposed model under various traffic profile and network configurations. The empirical results show that the proposed power consumption model can achieve an estimation error within 5\%, which is the lowest to the best of the author's knowledge.
  \item We defined three coverage scenarios for \cgls{nbiot} and \cgls{ltem} based on the coupling loss with the aim to compare the two technologies fairly. We also analyzed the impact of traffic profile, coverage scenario, and network configuration parameters on the device's battery lifetime based on the proposed model, shedding some light on how the \cgls{iot} application developers as well as the network operators should carefully configure the parameters to best tradeoff between the performance and device lifetime of their use case.
\end{enumerate}

The rest of the paper is structured as follows. A review of the latest literature on the power consumption of \cgls{iot} devices is presented in Section~\ref{sec:relatedwork}. Section~\ref{sec:state} introduces the \cgls{ue} states and procedures, followed by the energy consumption of those states and procedures in Section~\ref{sec:stateModel} and \ref{sec:procedureModel}, respectively. The considered coverage scenario, traffic profile, and the battery lifetime estimation model are presented in Section~\ref{sec:estimation}. The testbed setup and the measurement results are presented and discussed in Section~\ref{sec:results}. Conclusions are drawn in Section~\ref{sec:conclusion}. The main acronyms used in this paper are listed in \autoref{tab:acronyms}.

\begin{table}[t]
    \centering
	\begin{tabular}{c|l}
	\hline
	 \textbf{Acronym} & \textbf{Description} \\
	 \hline
	 \textbf{cDRX} & Connected mode Discontinuous Reception \\
	 \textbf{CE} & Coverage Enhancement \\
	 \textbf{CIoT} & Cellular Internet of Things \\
	 \textbf{CP} & Control Plane optimization \\
	 \textbf{DCI} & Downlink Control Information \\
	 \textbf{DL} & Downlink \\
	 \textbf{DRX} & Discontinuous Reception \\
	 \textbf{DUT} & Device Under Test \\
	 \textbf{ECM} & EPS Connection Management \\
	 \textbf{eDRX} & Extended Discontinuous Reception \\
	 \textbf{EMM} & EPS Mobility Management \\
	 \textbf{EPS} & Evolved Packet System \\
	 \textbf{iDRX} & Idle mode Discontinuous Reception \\
	 \textbf{IoT} & Internet of Things \\
	 \textbf{LPWAN} & Low Power Wide Area Network \\
	 \textbf{LTE} & Long Term Evolution \\
	 \textbf{LTE-M} & Long Term Evolution for Machines \\
	 \textbf{MCL} & Maximum Coupling Loss \\
	 \textbf{MCS} & Modulation and Coding Scheme \\
	 \textbf{NAS} & Non-Access Stratrum \\
	 \textbf{NB-IoT} & Narrowband Internet of Things \\
	 \textbf{NPDCCH} & Narrowband Physical Downlink Control Channel \\
	 \textbf{NPRACH} & Narrowband Physical Random Access Channel \\
	 \textbf{NPUCCH} & Narrowband Physical Uplink Control Channel \\
	 \textbf{NPUSCH} & Narrowband Physical Uplink Shared Channel \\
	 \textbf{OFDM} & Orthogonal Frequency-Division Multiplexing \\
	 \textbf{OFDMA} & Orthogonal Frequency-Division Multiple  Access \\
	 \textbf{PDCCH} & Physical Downlink Control Channel \\
	 \textbf{PRACH} & Physical Random Access Channel \\
	 \textbf{PRB} & Physical Resource Block \\
	 \textbf{PSM} & Power Saving Mode \\
	 \textbf{PTW} & Paging Time Window \\
	 \textbf{PUCCH} & Physical Uplink Control Channel \\
	 \textbf{PUSCH} & Physical Uplink Shared Channel \\
	 \textbf{RA} & Random Access \\
	 \textbf{RAP} & Random Access Preamble \\
	 \textbf{RAR} & Random Access Response \\
	 \textbf{RRC} & Radio Resource Control \\
	 \textbf{RU} & Resource Unit \\
	 \textbf{SC-FDMA} & Single-Carrier Frequency-Division  Multiple Access \\
	 \textbf{SF} & Subframe \\
	 \textbf{SR} & Scheduling Request \\
	 \textbf{TAP} & Test Automation Platform \\
	 \textbf{TAU} & Tracking Area Update \\
	 \textbf{TBS} & Transport Block Size \\
	 \textbf{UE} & User Equipment \\
	 \textbf{UL} & Uplink \\
	 \textbf{USS} & UE specific Search Space \\
	 \hline
	\end{tabular}
	\caption{List of Acronyms}
	\label{tab:acronyms}
\end{table}

\section{Related Work}\label{sec:relatedwork}
The power consumption model for regular broadband \cgls{lte} network has been addressed widely in the literature, as discussed in~\cite{mads} and the references therein. In recent years, various works have been published focusing on the power consumption analysis of \cgls{iot} devices via simulations, analytical modelling, or experimental measurements.

El Soussi \textit{et al}. \cite{Soussi} evaluated the performance of \cgls{nbiot} and \cgls{ltem} in the context of smart city use case, by using NS-3 simulation in an urban cell. It was found that an 8-year battery lifetime is achievable for both both technologies in a poor coverage scenario with a reporting interval of one day. The network coverage and capacity of various \gls{iot} technologies, including \gls{nbiot} and \gls{ltem}, have been evaluated in \cite{ltemnbiotcovcap,Vejlgaard} through both simulation and empirical measurements. It was found that \gls{nbiot} provides better coverage, especially for indoor cases, at the cost of lower capacity in terms of the total number of supported devices. The authors also showed the results of the average power consumption for the two technologies in various propagation scenarios. In addition to their previous work, the power consumption and a battery lifetime estimation model of two \cgls{nbiot} devices were proposed in \cite{GermanPaper} using experimental measurements. However, the abstraction level of the power consumption model is relatively high, which reduces the complexity of the model, but also introduces inaccuracies and obfuscates energy trade-offs at lower-level mechanisms. An energy consumption model for \gls{nbiot} devices considering \cgls{psm} and \cgls{edrx} with different timer parameters was proposed in \cite{nbiotpsmedrx}. The analysis is compared to the NS-3 simulation results, showing an average inaccuracy of around 11.8~\%. A tractable theoretical model of energy consumption in \cgls{nbiot} was presented in~\cite{Azari}, providing some insights in the tradeoff between energy efficiency, latency and coverage. Andres-Maldonado \textit{et al}. \cite{anamodnbiot,nbiot} proposed a Markov chain based analytical model to estimate the energy consumption and delay of a \cgls{nbiot} device sending periodic uplink reports using the control plane procedure. The model has been validated using two commercial \cgls{nbiot} device and three test cases, obtaining a maximum relative error of the battery lifetime estimation of 21\%. It is noted that the periodic \cgls{tau} procedure is not considered in \cite{anamodnbiot}.

Hertlein \textit{et al}. \cite{Hertlein} compared the power consumption of several cellular standards via measurements. It is confirmed that \cgls{nbiot} can reduce the power consumption as compared to other cellular standards. Experimental assessment of the expected lifetime of an \gls{nbiot} device is presented in \cite{Duhovnikov}, based on measurements from both a commercial and a private network in the context of aviation use case. It is found that the battery lifetime can be extended significantly by using \gls{psm} in the idle state. Although the results obtained in \cite{Duhovnikov} have high credibility in terms of realism, it is not straightforward to extrapolate those results to different network scenarios, traffic patterns, and transmission parameter configurations. An End-to-End (E2E) integration study of an experimental \cgls{nbiot} trial network was perform in \cite{Yeoh} by Telekom Malaysia. The assessment results showed that a careful setup of the \cgls{nbiot} device’s firmware is critical to prolong the battery lifetime, e.g., avoiding unnecessary network attach and detach procedures, and proper configuration of the T3324 timer and T3412 timer to best match the tradeoff between delay and energy efficiency, etc. Martinez \textit{et al}. \cite{Martinez} empirically investigated the performance boundaries of \cgls{nbiot} in terms of energy consumption, reliability, and delays. The measurements were performed over a commercial  \cgls{nbiot} network in Spain with two commercial \cgls{nbiot} devices, using different device and network configurations. It is found that some of the timers associated with the power saving mechanisms have an obvious effect on the device energy consumption. Abbas \textit{et al}. \cite{Abbas} proposed guidelines on how to configure both tunable and non-tunable parameters with the target to conserve the energy consumption by means of simulations. A more comprehensive measurement study of the energy consumption is conducted in \cite{Michelinakis} with two popular \cgls{nbiot} devices in two major western European operators under different parameter configurations, with the aim to identify the key parameters for enhancing the battery lifetime. Its findings indicate that careful configurations are required to improve the energy efficiency. For higher level \cgls{iot} applications, Lin \textit{et al}. \cite{Lin} developed a service platform for fast development of \cgls{nbiot} applications, which is used in a smart parking application as an example.

The main issue we observed in the current state of the art work on analytical energy consumption modelling of \cgls{iot} devices is the lack of accuracy and flexibility. Besides, although quite much work has been done trying to fine tune the parameters to lower the energy consumption, those results are mainly obtained from extensive experimental measurements. The main objective of this work is to first propose a highly accurate framework for modelling the power consumption of \gls{nbiot} and \gls{ltem}, consisting of detailed modelling of \gls{ue} states and procedures. The model is flexible for different configurations of traffic profile, transmission and network parameters. Based on the proposed model, we also attempt to identify the parameters that mostly impact the battery lifetime. Our findings are in line with the observations obtained from the measurements as mentioned in the previous literature.

%% file: 02_PowerModel.tex
\section{UE States and Procedures}\label{sec:state}
Our framework for the power consumption modelling is based on the states of the \cgls{ue} and the transition among the states which is triggered by the procedures. The main differences between \cgls{nbiot} and \cgls{ltem} will be described first, followed by a general description of the \cgls{ue} states and procedures.

\subsection{Differences between \cgls{nbiot} and \cgls{ltem} \label{subsec:difference}}
Both \cgls{ltem} and \cgls{nbiot} are developed based on \cgls{lte} with the aim to offer better coverage and lower energy consumption. Although these two technologies share a lot of similarities, they are targeted for different use cases and therefore have some differences. Comparatively, \cgls{ltem} supports higher data rate, lower latency, and higher mobility, while \cgls{nbiot} is focused on enhanced coverage, longer battery life, and lower device complexity \cite{Ghosh}.

\cgls{ltem} has defined two \cgls{ce} Modes: Mode A and Mode B, targeting for different coverage levels \cite{coverageAnalysisLTEM}. This paper only considers \cgls{ce} Mode A, as this is a mandatory feature for \cgls{ltem} while Mode B is optional. Some of the main differences between the two technologies are summarized in \autoref{tab:difference}. Specifically, the transmission bandwidth of \cgls{nbiot} is reduced from 1.08 MHz to 180 kHz, resulting in a lower data rate as compared to \cgls{ltem}. In downlink, both \cgls{ltem} and \cgls{nbiot} support multi-tone transmissions based on \cgls{ofdma} with 12 subcarriers and 15 kHz subcarrier spacing. In uplink, \cgls{ltem} only supports  multi-tone transmissions based on \cgls{scfdma} while \cgls{nbiot} supports both single-tone and multi-tone transmissions. The minimum resource allocation unit in both uplink and downlink for \cgls{ltem} is a \cgls{prb}, consisting of 12 subcarriers and 1 slot (7 OFDM symbols). The minimum resource allocation unit in downlink for \cgls{nbiot} is also a \cgls{prb}, while in uplink the minimum unit is a \cgls{ru}, of which the number of subcarriers and slots within a \cgls{ru} depend on the transmission option (i.e., single-tone or multi-tone) \cite{Kanj}. The number of repetitions in \cgls{nbiot} has increased up to 128 in uplink and 2048 in downlink to further extend the coverage level. By trading off data rate and latency, i.e., reducing the transmission bandwidth and/or increasing the number of retransmissions, the coverage in terms of \cgls{mcl} is extended to 164 dB in \cgls{nbiot} (20 dB deeper than \cgls{lte}).

As previously stated, the physical layer channels and signals of both \cgls{ltem} and \cgls{nbiot} are mostly inherited from \cgls{lte}, but are optimized to meet the constraints and requirements of \cgls{iot} modules. There is not much difference for the downlink channels between \cgls{ltem} and \cgls{nbiot}. However in the uplink, \cgls{nbiot} transmits the data and control information both in \cgls{npusch} with different formats \cite{Kanj}, while \cgls{ltem} transmits data in \cgls{pusch} and control information in \cgls{pucch}.

\begin{table}[t]
    \centering
    %\begin{tabularx}{|c|c|c|}
    \begin{tabularx}{\linewidth}{|X|p{3cm}|p{3cm}|}
		\hline
		\textbf{Attribute} & \textbf{\glsentryshort{ltem}} & \textbf{\glsentryshort{nbiot}} \\ \hline
		Deployment & In-band & In-band, Guard-band, Stand-alone \\ \hline
		Coverage  & 155.7 dB \cgls{mcl} & 164 dB \cgls{mcl} \\ \hline
		Bandwidth  & 1.08 MHz & 180 kHz \\ \hline
		Downlink Transmission & \cgls{ofdma}, multi-tone with 12 subcarriers, 15 kHz tone spacing & \cgls{ofdma}, multi-tone with 12 subcarriers, 15 kHz tone spacing \\ \hline
		Uplink Transmission & \cgls{scfdma}, multi-tone with 12 subcarriers, 15 kHz tone spacing & Multi-tone with 3, 6 or 12 subcarriers (\cgls{scfdma}), 15 kHz tone spacing \\
		  &  & Single-tone, 3.75 kHz \& 15 kHz tone spacing  \\ \hline
		Resource & \cgls{prb} in DL & \cgls{prb} in DL \\ 
		Allocation & \cgls{prb} in UL & \cgls{ru} in UL \\ \hline
		Max number & 32 in UL (\cgls{ce} mode A) & 128 in UL \\
		of repetitions & 32 in DL (\cgls{ce} mode A) & 2048 in DL \\ \hline
		Peak rate  & 1 Mbps for DL and UL & DL: 250 Kbps \\
		&  & UL: 250 Kbps(multi-tone) \\
		&  & UL: 20 Kbps (single-tone) \\ \hline
		Uplink Channel & \cgls{pucch}: Transmission of control information & \cgls{npusch} format 2: Transmission of control info \\
		 & \cgls{pusch}: Transmission of data & \cgls{npusch} format 1: Transmission of data \\
		 & \cgls{prach}: Transmission of preambles & \cgls{nprach}: Transmission of preambles with a new waveform \\ \hline
	\end{tabularx}
    \caption{Differences between \cgls{nbiot} and \cgls{ltem}.}
    \label{tab:difference}
\end{table}

\subsection{UE Procedures}\label{subsec:procedures}
To be able to establish a connection to the network and start data transmission, the \cgls{ue} has to perform a series of actions with specific purposes and functions, such as frequency and time synchronization, performing \cgls{ra}, sending service request, etc. These actions can be referred to as procedures. Some of the main procedures performed in \cgls{nbiot} and \cgls{ltem} are listed below:

\begin{itemize}
    \item \textbf{Synchronization:} When the \cgls{ue} is switched on or wakes up, it has to synchronize with the network both in time and frequency domains by decoding synchronization signals, namely \cgls{pss} and \cgls{sss}. After the completion of synchronization and cell search, the \cgls{ue} can perform the \cgls{ra} procedure.
    \item \textbf{Attach:} If the \cgls{ue} has the intention to communicate with the network but has not been registered in the network yet, i.e., in \glss{emmd} state, it performs the attach procedure after being synchronized to the network. The attach procedure is used to create the \cgls{eps} bear between the \cgls{ue} and the \cgls{pgw} in order to be able to send and receive data.
    \item \textbf{Service Request:} The service request procedure is similar to the attach procedure, but with fewer steps. It is used to activate a user plane connection for data transmission when the \cgls{ue} has already been registered in the network, i.e., in \glss{emmr} state. It will bring the \cgls{ue} from \glss{ecmi} state into \glss{ecmc} state.
    \item \textbf{Connection Resume:} The \cgls{rrc} connection resume procedure is initiated by the \cgls{ue} when upper layers requests the resume of a suspended \cgls{rrc} connection. It is used as a replacement of the service request procedure if the connection has been suspended instead of released. The \cgls{ue} should have valid and up to date essential system information before initiating the connection resume procedure.
    \item \textbf{Connection Release:} The release procedure is used to release an \cgls{rrc} connection. If the \cgls{ue} does not have any activity in the network but still would like to be registered in the network, the network initiates the \cgls{rrc} connection release procedure to the \cgls{ue}. The \cgls{ue} can still be contacted via paging after the release procedure. If the \cgls{ue} has data to transmit, it has to perform a service request procedure.
    \item \textbf{\cgls{tau}:} A \cgls{ue} in \glss{ecmi} state sends a \cgls{tau} message to the \cgls{mme} periodically to notify to the network that it is still alive and is able to receive data. Otherwise, the network would assume the \cgls{ue} is not reachable and will not perform paging even when there is data traffic for it.
\end{itemize}

The power consumption modelling of the aforementioned procedures will be described in Section~\ref{sec:procedureModel}.

\subsection{UE States}\label{subsec:states}
The \cgls{ue} states can be defined in several aspects: from a device perspective and from a network perspective. 

\begin{figure}[ht]
    \centering
    \includegraphics[width=\linewidth]{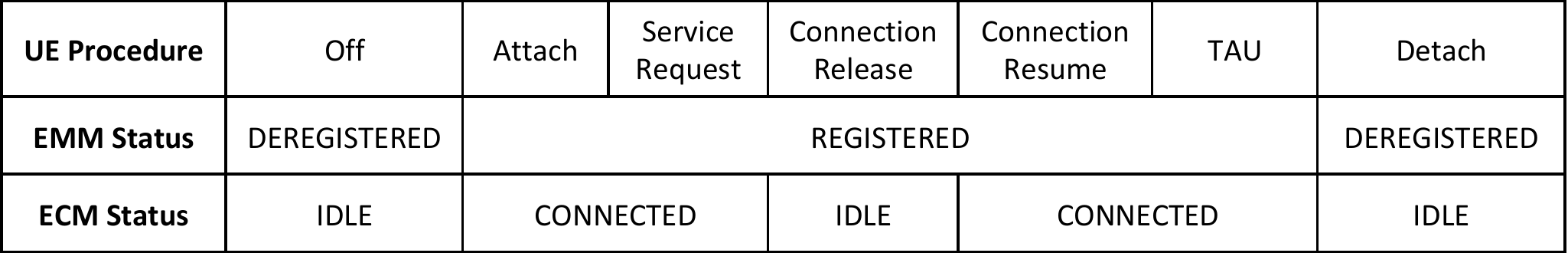}
    \caption{Example of UE procedures and the corresponding EMM/ECM network status.}
    \label{fig:EMMstatus}
\end{figure}

From a network perspective, the network status of the \cgls{ue} indicates how reachable the \cgls{ue} is and what are the available resources for the \cgls{ue}. It is the \cgls{emm} and \cgls{ecm} protocols that determine the network status of the \cgls{ue}. An example of \cgls{ue} procedures and the corresponding network status is illustrated in \autoref{fig:EMMstatus}.

The \cgls{emm} protocol indicates whether the \cgls{ue} is registered in the network or not. After the \cgls{ue} is switched on, it performs the synchronization and cell search procedure before entering the Synchronized state. At this moment, the \cgls{ue} is in \glss{emmd} state. The \cgls{emm} state of the \cgls{ue} is changed to \glss{emmr} after the \cgls{ue} has performed a network attach procedure, i.e., \cgls{rrc} Connection setup including the \cgls{ra} procedure, \cgls{nas} authentication and  \cgls{nas} security. The \cgls{emm} state of the \cgls{ue} is changed back to \glss{emmd} when either the \cgls{tau} timer has expired or a detach procedure is performed. 

The \cgls{ecm} protocol indicates whether the \cgls{ue} has established signalling to the \cgls{epc} or not. In the \glss{ecmc} state, the \cgls{ue} has an active \cgls{rrc} connection with the network and is semi-actively monitoring the network in order to preserve energy. In the \glss{ecmi} state, there is no active \cgls{rrc} connection and the \cgls{ue} will only be reachable in certain time intervals according to the selected power saving technique, which can be for example \cgls{drx} and \cgls{psm}. There are two types of \cgls{drx}. When the \cgls{ue} is in \glss{ecmc} state it is named as \cgls{cdrx} and when the \cgls{ue} is in \glss{ecmi} state it is named as \cgls{idrx}. The \cgls{ue} changes from \glss{ecmi} to \glss{ecmc} state by performing either a service request, an attach, or a connection resume procedure. To change back from \glss{ecmc} to \glss{ecmi}, the \cgls{ue} performs either a detach, a connection suspend, or a release procedure. 

\begin{figure}[tb]
    \centering
    \includegraphics[width=\linewidth]{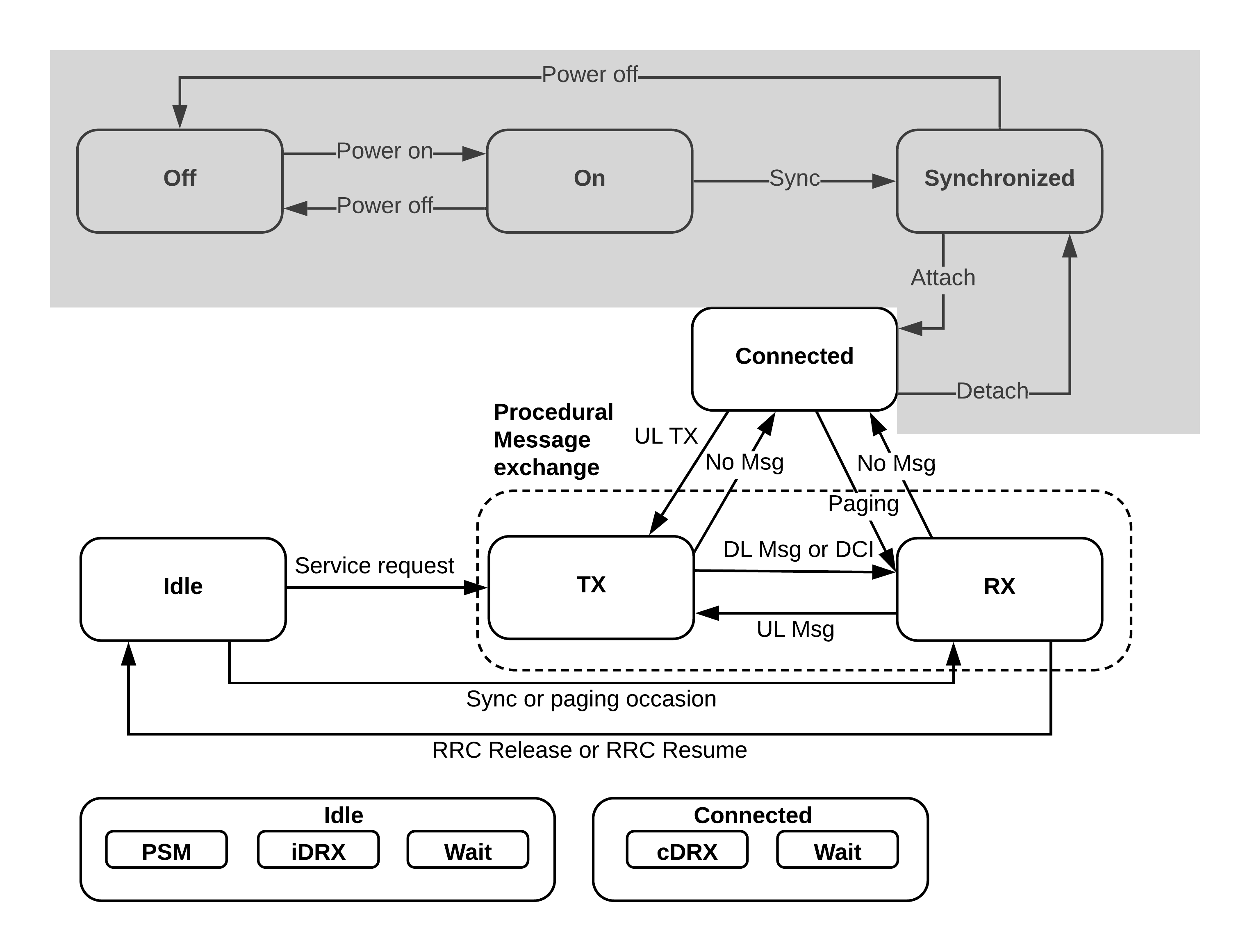}
    \caption{The \cgls{ue} state machine used for \cgls{ltem} and \cgls{nbiot} \cite{iotbook}.
    %\todo[inline]{RBS: corrections for this figure: IDLE should be connected to RX by PAGING and either SYNC or SERVICE REQUEST. ECM\_CONNECTED$->$RX should be PAGING. ECM\_CONNECTED$->$TX should be SERVICE REQUEST. The RELEASE and SERVICE REQUEST procedures should each only connect a single state to another state. I'd say IDLE to RX for SERVICE REQUEST and TX to IDLE for release. Finally the state cDRX/eDRX should be ECM\_CONNECTED and Idle should just be IDLE.}}
    %\todo[inline, author=Nicolaj]{Figure is created with lucidchart (link: https://lucid.app/lucidchart/invitations/accept/c5447708-af0c-421f-b1cd-a5777a7dbabb)
    }
    \label{fig:StateModel}
\end{figure}

From a device perspective, the \cgls{ue} can be in different states such as \cgls{rx}, \cgls{tx}, idle, etc. The state diagram for the \cgls{ue} is depicted in \autoref{fig:StateModel} together with the procedures required to move from one state to another. The grey area indicates the \cgls{ue} is in \glss{emmd} while the white area indicates the \cgls{ue} is in \glss{emmr}.

There is a power cost associated with \cgls{ue} state transitions in addition to the power cost of staying in a specific \cgls{ue} state. By characterising the power consumption of each \cgls{ue} state and the associated procedure(s), the power consumption of any \cgls{ue} behaviour can be modelled by composing different components of the states and procedures illustrated in \autoref{fig:StateModel}. The modelling of power consumption for different \cgls{ue} states and procedures will be presented in the following sections.

\section{Energy Consumption of \cgls{ue} States} \label{sec:stateModel}
In this section, the energy consumption models for different \cgls{ue} states are presented. For brevity, the details of the performed power measurements in each \cgls{ue} state are not included in this paper, but are available in \cite{iotbook}.

\subsection{\cgls{tx} State} \label{sec:state_tx}
In \cgls{tx} state, the \cgls{ue} is transmitting data in the uplink. The energy consumption in the \cgls{tx} state is determined by the \cgls{ue}'s power level as well as the length of the transmission. The \cgls{ue}'s required transmission power is controlled by the uplink power control specified in \cite{TS36213}. From measurements, it is observed that the \cgls{ue}'s uplink transmission power is independent of the selection of \cgls{mcs}.

\subsubsection{Uplink Packet}
The uplink data packets are transmitted in \cgls{pusch} and \cgls{npusch} with format 1 for \cgls{ltem} and \cgls{nbiot}, respectively. During uplink transmission, a mandatory gap of 40 milliseconds after 256 milliseconds of continuous transmission is introduced to allow the low-quality oscillators to resynchronize with the network \cite{TS36211,TS36331}. The power consumption is much lower during these transmission gaps as compared to the power consumption during data transmission. 

The estimated energy consumption for the \cgls{tx} state with an uplink packet of payload size $k$ can be calculated as:
\begin{equation}\label{eq:TX_energy}
    E_{\text{TX}}(k) = t_{\text{TX}}(k)\cdot P_{\text{TX}} + t_{\text{TX}}^{\text{gaps}}(t_\text{TX})\cdot P_{\text{TX}}^{\text{gaps}}
\end{equation}
where $t_{\text{TX}}(k)$ is the time spent in data transmission with $k$ bits, $P_{\text{TX}}$ is the power consumption during data transmission, $t_{\text{TX}}^{\text{gaps}}(t_\text{TX})$ is the time spent in transmission gaps, and $P_{\text{TX}}^{\text{gaps}}$ is the power consumption during transmission gaps. 

The transmission time $t_{\text{TX}}(k)$ for a packet with payload size $k$ in bits depends on the selection of \cgls{mcs}, the number of allocated \cgls{ru}s or \cgls{sf}s, and the number of repetitions. For \glss{nbiot} and \glss{ltem}, the transmission time can be calculated as:.
\begin{align}
    t_{\text{TX}}^{\text{NB-IoT}}(k) &= t_{\text{RU}} \cdot \rho_{\text{RU}} \cdot \text{N}_\text{pkt}^{\text{NB-IoT}} \cdot l^{\text{NB-IoT}}(k) \label{eq:tTXNBiot}\\
    l^{\text{NB-IoT}}(k) &= \ceil*{\frac{k}{TBS(MCS,\rho_{\text{RU}})- h^{\text{NB-IoT}}}} \\
    t_{\text{TX}}^{\text{LTE-M}}(k) &= t_{\text{SF}} \cdot \rho_{\text{SF}} \cdot \text{N}_\text{pkt}^{\text{LTE-M}} \cdot l^{\text{LTE-M}}(k) \label{eq:tTXLTEM}\\
    l^{\text{LTE-M}}(k) &= \ceil*{\frac{k}{TBS(MCS,\rho_{\text{SF}})- h^{\text{LTE-M}}}}
\end{align}
where $t_{\text{RU}}$ is the duration of a \cgls{ru}, $\rho_{\text{RU}}$ is the number of allocated \cgls{ru}s, $t_{\text{SF}}$ is the duration of a \cgls{sf}, $\rho_{\text{SF}}$ is the number of allocated \cgls{sf}s, $l^{\text{NB-IoT}}(k)$ and $l^{\text{LTE-M}}(k)$ are the number of segments, $\text{N}_\text{pkt}^{\text{NB-IoT}}$ and $\text{N}_\text{pkt}^{\text{LTE-M}}$ are the number of repetitions, TBS is the transport block size determined by the selected \cgls{mcs} and the number of allocated resources in terms of \cgls{ru}(s) for \cgls{nbiot} and \cgls{sf}(s) for \cgls{ltem},  $h^{\text{NB-IoT}}$ and $h^{\text{LTE-M}}$ are the header size, and $\ceil{}$ is the ceil function. The length of a \cgls{sf} ($t_{\text{SF}}$) is always 1 ms, while the length of a \cgls{ru} ($t_{\text{RU}}$) can range from 1 ms to 32 ms, depending on the configured uplink transmission option, e.g., subcarrier spacing, the number of configured subcarriers and slots. The available transmission options of one \cgls{ru} can be referred to \cite{Kanj}. It should be noted that \autoref{eq:tTXLTEM} does not hold if frequency hopping is enabled for \cgls{ltem}. In that case, \cgls{prb}s are scheduled over multiple \cgls{sf}s.

The time spent in transmission gaps depends on $t_{\text{TX}}$, and can be calculated as:
\begin{equation}
    t_{\text{TX}}^{\text{gaps}}(t_\text{TX}) = \floor*{\frac{t_{\text{TX}}}{t_{\text{TX}_{\text{max}}}}} \cdot t_{\text{TX}}^{\text{gap}} \label{eq:TXGapsModelTime}
\end{equation}
where $t_{\text{TX}_{\text{max}}}$ is the maximum continuous transmission time allowed, $t_{\text{TX}}^{\text{gap}}$ is the duration of a gap, and $\floor{}$ is the floor function. According to \cite{TS36211}, $t_{\text{TX}_{\text{max}}}$ and $t_{\text{TX}}^{\text{gap}}$ for NB-IoT are 256 ms and 40 ms, respectively.

\subsubsection{Uplink Control}
Uplink control information (e.g., ACK/NACK) is transmitted in \cgls{pucch} for \cgls{ltem} and in \cgls{npusch} with format 2 for \cgls{nbiot}, respectively. It should be noted that using \cgls{npusch} format 2, the \cgls{ru} is always configured with one subcarrier and 4 slots. The energy consumption for transmitting uplink control information can be calculated similarly as for uplink data packet:

\begin{align}
    E_{\text{TX(ack)}} &= t_{\text{TX(ack)}}\cdot P_{\text{TX}} + t_{\text{TX(ack)}}^{\text{gaps}}\cdot P_{\text{TX}}^{\text{gaps}} \\
    t_{\text{TX(ack)}}^{\text{NB-IoT}} &= t_{\text{RU(ack)}} \cdot \text{N}_\text{ack}^{\text{NB-IoT}} \\
    t_{\text{TX(ack)}}^{\text{LTE-M}} &= t_{\text{SF}} \cdot \text{N}_\text{ack}^{\text{LTE-M}} \\
    t_{\text{TX(ack)}}^{\text{gaps}} &= \floor*{\frac{t_{\text{TX(ack)}}}{t_{\text{TX}_{\text{max}}}}} \cdot t_{\text{TX}}^{\text{gap}}
\end{align}

\subsubsection{Random Access}
The transmission of preamble is performed in \cgls{prach} for \cgls{ltem} and in \cgls{nprach} for \cgls{nbiot}, respectively. There is no transmission gap during the transmission in \cgls{rap}. So the energy consumption for transmitting the preamble can be calculated as:
\begin{align} \label{eq:RAPenergyconsumption}
    E_{\text{TX}}^{\text{RAP}} = t_{\text{RAP}} \cdot P_{\text{TX}}
\end{align}
where $t_{\text{RAP}}$ is the duration of the \cgls{rap}, and can be calculated as:
\begin{align}
    t_{\text{RAP}}^\text{NB-IoT}& =  \left( t_{\text{CP}}^{\text{format}} + 5  \cdot t_{\text{RASym}} \right) \cdot 4 \cdot \text{N}_{\text{RAP}} \nonumber\\
    t_{\text{RAP}}^\text{LTE-M}& =  t_{\text{RAP}}^{\text{format}} \cdot \text{N}_{\text{RAP}}
\end{align}
where $t_{\text{CP}}^{\text{format}}$ is the duration of the cyclic prefix, $ t_{\text{RASym}} $ is the symbol duration in \cgls{nbiot} \cgls{nprach}, $t_{\text{RAP}}^{\text{format}}$ is the duration of the \cgls{rap} for \cgls{ltem}, and $\text{N}_{\text{RAP}}$ is the number of repetitions for \cgls{rap}. For \cgls{nprach} in \glss{nbiot}, $t_{\text{RASym}}$ is equal to 266.7 $\mu s$ (3.75 kHz subcarrier spacing), $t_{\text{CP}}^{\text{format}}$ is equal to 66.7 $\mu s$ for format 0 and 266.7 $\mu s$ for format 1. The specific value of $t_{\text{RAP}}^{\text{format}}$ can be referred to Table 5.7.1-1 in \cite{TS36211}, e.g., \SI{0.903}{\milli\second} for format 1.

\subsection{\cgls{rx} State} \label{sec:state_rx}
In \cgls{rx} state, the \cgls{ue} is receiving data in the downlink. From the measurement, it is observed that the \cgls{rx} state is characterized by periods of reception interleaved by reception gaps. \cgls{ltem} and \cgls{nbiot} have a similar downlink channel, and can be modelled in a similar way. The estimated energy consumption of the \cgls{rx} state with a downlink packet of size $k$ in bits can be calculated as:
\begin{align}
    E_{\text{RX}}(k) &= P_{\text{RX}} \cdot t_{\text{RX}}(k) + P_{\text{RX}}^{\text{gaps}} \cdot t_{\text{RX}}^{\text{gaps}}(t_{\text{RX}})  \label{eq:RxState} \\
    t_{\text{RX}}(k) &= t_{\text{SF}} \cdot \rho_{\text{SF}} \cdot \text{N}_{\text{SF}} \cdot l(k) \label{eq:SFTime} \\
    l(k) &= \ceil*{\frac{k}{TBS(MCS,\rho_{\text{SF}})- h}} \label{eq:segment}
\end{align}
where $t_{\text{RX}}$ and $P_{\text{RX}}$ indicate the time spent and the power consumption in data reception respectively, $t_{\text{RX}}^{\text{gaps}}$ and $P_{\text{RX}}^{\text{gaps}}$ indicate the time spent and the power consumption in reception gaps respectively, $t_{\text{SF}}$ is the length of a subframe which is 1 ms in both \cgls{nbiot} and \cgls{ltem}, $\rho_{\text{SF}}$ is the number of allocated \cgls{sf}s, $\text{N}_{\text{SF}}$ is the number of repetitions, $l$ is the number of segments, and $h$ is the header size.

Reception gaps are introduced in the downlink channel because of the reception of \cgls{sib} and other control signalling \cite{TS36213,TS36211}. Similar to the \cgls{tx} state, the power consumption is much lower during the reception gaps as compared to the power consumption during data reception. Also the power consumption in \cgls{rx} state is \cgls{mcs} independent. Different from the \cgls{tx} model where the transmission gaps follow a clear pattern, it is not straightforward to accurately model the reception gaps in \cgls{rx} state because they are dependent on the length of the reception as well as its starting position. Instead, the number of reception gaps during a data reception is estimated. The time spent in reception gaps is dependent on $t_{\text{RX}}$ and can be calculated as:
\begin{align}
	t_{\text{RX}}^{\text{gaps}}(t_{\text{RX}}) &= \ceil*{t_{\text{RX}} \cdot \left( \frac{1}{\text{M}_{\text{SF}}^{\text{av}}} - 1 \right)} \label{eq:SFGap}
\end{align}
where $\text{M}_{\text{SF}}^{\text{av}}$ is the fraction of subframes available for data reception. To simplify the model, only the most recurring gaps such as \cgls{npss}, \cgls{nsss}, \cgls{mib}, and \cgls{sib}1 are taken into account when modelling the \cgls{rx} state, resulting in roughly 14 subframes out of 20 subframes are available for downlink data reception ($\text{M}_{\text{SF}}^{\text{av}}=14/20$) \cite{TS36211,TS36331}.

\subsection{ECM Connected State} \label{sec:state_ecmc}
A \cgls{ue} that is not transmitting or receiving data in the \glss{ecmc} state must monitor the network for paging or uplink scheduling grants in the form of \cgls{dci}. This can happen in default or \cgls{cdrx} mode, as described in the following.

\subsubsection{Default mode} \label{sec:defmod}
By default the \cgls{ue} monitors the \cgls{uss} in \cgls{pdcch} for \cgls{ltem} or in \cgls{npdcch} for \cgls{nbiot} continuously for relevant \cgls{dci}, i.e., paging and uplink grants. The \cgls{uss} must be monitored periodically and the \cgls{ue} will go into sleep for the rest of the time. The time duration for the \cgls{uss} monitoring in each cycle is given by:
\begin{align}\label{eq:PeriodUSS}
    t_{\text{USS}}^{\text{monitoring}} &= \text{M}_{\text{USS}} \cdot t_{\text{SF}} \cdot \text{N}_{\text{USS}}
\end{align}
where $\text{M}_{\text{USS}}$ is the number of subframes that the \cgls{ue} has to monitor in \cgls{pdcch} or \cgls{npdcch}, and $\text{N}_{\text{USS}}$ is the number of repetitions for \cgls{uss}.  

The estimated energy spent in each cycle using default \cgls{uss} monitoring in the \glss{ecmc} state can be calculated as:
\begin{align}
    E_{\text{ECMC}}^\text{cycle} &= E_{\text{USS}}^{\text{monitoring}} +  P_{\text{USS}}^{\text{sleep}} \cdot t_{\text{USS}}^{\text{sleep}} \label{eq:USSEnergy}
\end{align}
where $E_{\text{USS}}^{\text{monitoring}}$ is the energy consumption for \cgls{uss} monitoring which can be obtained using \autoref{eq:RxState,eq:SFGap} with $t_{\text{USS}}^{\text{monitoring}}$ as input.

\subsubsection{\gls{cdrx}} \label{sec:cdrx}
\cgls{drx} allows the \cgls{ue} to monitor the relevant \cgls{uss} discontinuously to improve the energy efficiency. While \cgls{idrx} operates in the \glss{ecmi} state, \cgls{cdrx} operates in the \glss{ecmc} state. In connected state the \cgls{ue} does not perform paging, but rather monitors the \cgls{pdcch} or \cgls{npdcch}. \cgls{cdrx} essentially provides a way for the network and the \cgls{ue} to synchronize the timing of potential downlink data. It narrows down the \cgls{uss} window, which enables the \cgls{ue} to sleep for much longer periods. The \cgls{ue} notifies the network of its non-mandatory \cgls{drx} capabilities through \cgls{rrc} messaging and the network transmits the \cgls{drx} configuration as part of the \cgls{rrc} configuration.

The synchronization between the network and \cgls{ue} in \cgls{drx} mode is controlled by the LongDRX-cyle and the OnDuration timer \cite{TS36331}. The OnDuration timer sets the number of consecutive \cgls{pdcch} subframes that the \cgls{ue} has to monitor for \cgls{dci}. The value of the OnDuration timer can be set up to  \SI{200}{\milli\second}. The time that the \cgls{ue} has to monitor for \cgls{dci} in the OnDuration is given by:
\begin{align}
    t_{\text{cDRX}}^{\text{onDur}} &= \text{M}_\text{onDur} \cdot t_{\text{SF}} \label{eq:OnDurationTime}
\end{align}
Assuming that the OnDuration timer stays constant during the \cgls{ue}'s lifetime, the energy consumption of the \cgls{cdrx} can be calculated as:
\begin{align} 
    E_{\text{cDRX}} &= E_{\text{onDur}} + P_{\text{cDRX}}^{\text{sleep}}\cdot t_{\text{cDRX}}^{\text{sleep}} \label{eq:cDRXEnergy} \\
    t_{\text{cDRX}}^{\text{sleep}} &=  (\text{M}_{\text{LongDRX}}^\text{cycle} - \text{M}_{\text{onDur}}) \cdot t_{\text{SF}} \label{eq:cDRXTime}
\end{align}
where $\text{M}_{\text{LongDRX}}^\text{cycle}$ and $\text{M}_{\text{onDur}}$ are the number of subframes of a \cgls{cdrx} LongDRX-cyle and the number of subframes the \cgls{ue} has to monitor for \cgls{dci} in the OnDuration respectively, $P_{\text{cDRX}}^{\text{sleep}}$ is the power consumption when the \cgls{ue} is in \cgls{cdrx} sleep mode, and $E_{\text{onDur}}$ is the energy consumption in the OnDuration which can be calculated using \autoref{eq:RxState,eq:SFGap} with $t_{\text{cDRX}}^{\text{onDur}}$ as input. It is worth mentioning that the inactivity timer which triggers the \cgls{ue} to enter into the \cgls{cdrx} state is neglected here,  because in most cases the \cgls{ue} is not expected to frequently alternate between \cgls{rx} and \cgls{cdrx} as it is taxing for the battery.

\subsection{ECM Idle State} \label{sec:state_ecmi}
A \cgls{ue} in the \glss{ecmi} state must monitor the network for paging for relevant \cgls{dci}. Several energy saving modes are available.

\subsubsection{Default mode}
By default the \cgls{ue} monitors the \cgls{uss} in \cgls{pdcch} or \cgls{npdcch} continuously for paging. The energy consumption in the \glss{ecmi} state operating in default mode is similar to the approach described in \autoref{sec:defmod}.

\subsubsection{\cgls{idrx}} \label{sec:drx}
\cgls{drx} was introduced in release 8 to save the energy consumption of a device by introducing the \cgls{drx} cycle, during which the \cgls{ue} alternates between active monitoring of the paging and sleep \cite{lte}. During the active period, the \cgls{ue} first gets synchronized with the network, then it monitors the paging. The number of paging occurrences the \cgls{ue} has to monitor is indicated by the paging repetitions \cite{TS36331}. When the \cgls{ue} has finished listening to the paging, it goes into sleep. The duration of a \cgls{drx} cycle is given by the network parameter defaultPagingCycle in number of radio frames.

The energy consumption of the \cgls{idrx} can be calculated as:
\begin{align}
    E_{\text{iDRX}}^{\text{cycle}} &= E_{\text{iDRX}}^{\text{sync}} + E_{\text{paging}} \cdot \text{N}_{\text{paging}} + P_{\text{iDRX}}^{\text{sleep}}\cdot t_{\text{iDRX}}^{\text{sleep}} \\
    t_{\text{iDRX}}^{\text{sleep}} &=  t_{\text{iDRX}}^{\text{cycle}} - ( t_{\text{iDRX}}^{\text{onDur}} + t_{\text{iDRX}}^{\text{sync}})
\end{align}
where $t_{\text{iDRX}}^{\text{sync}}$ and $E_{\text{iDRX}}^{\text{sync}}$ are the measured time and energy consumption for the synchronization in an \cgls{idrx} cycle, $t_{\text{iDRX}}^{\text{sleep}}$ and $P_{\text{iDRX}}^{\text{sleep}}$ are the time and power consumption in \cgls{idrx} sleep state, $E_{\text{paging}}$ is the energy spent on a single paging occasion, $\text{N}_{\text{paging}}$ is the number of paging repetitions in an \cgls{idrx} cycle, $t_{\text{iDRX}}^{\text{onDur}}$ is the time spent on monitoring the paging in an \cgls{idrx} cycle, and $t_{\text{iDRX}}^{\text{cycle}}$ is the length of an \cgls{idrx} cycle.

\subsubsection{\cgls{edrx}} \label{sec:edrx}
\cgls{edrx} is an extension of the \cgls{drx} with the objective to further reduce the energy consumption. It is mainly used in \cgls{iot} applications operating with energy saving mode. The basic principle of \cgls{edrx} is to extend \cgls{drx} cycle length to allow a device to remain in sleep mode for longer period of time, thus reducing the energy consumption. Furthermore, the length of an \cgls{edrx} cycle can be set by the device rather than the network, which provides the  application developers with more flexibility to better balance the device’s reachability and its energy consumption. An \cgls{edrx} cycle can be decomposed into two parts: the \cgls{ptw} period and the sleep period. During the \cgls{ptw} period, the \cgls{ue} behaves similarly as being in \cgls{idrx} state with several \cgls{idrx} cycles. The number of \cgls{idrx} cycles is determined by the \cgls{ptw} length and the \cgls{idrx} cycle length. The \cgls{ue} remains dormant during the sleep period, the duration of which can be calculated based on the \cgls{ptw} length and the \cgls{edrx} cycle length.

The energy consumption of the \cgls{edrx} can be calculated as:
\begin{align}
    E_{\text{eDRX}}^{\text{cycle}} &=  \ceil[\bigg]{\frac{T_{\text{PTW}}}{t_{\text{iDRX}}^{\text{cycle}}}} \cdot E_{\text{iDRX}}^{\text{cycle}} + t_{\text{eDRX}}^{\text{sleep}} \cdot P_{\text{eDRX}}^{\text{sleep}} \\
    t_{\text{eDRX}}^{\text{sleep}} &= t_{\text{eDRX}}^{\text{cycle}} - T_{\text{PTW}} 
\end{align}
where $t_{\text{eDRX}}^{\text{sleep}}$ and $P_{\text{eDRX}}^{\text{sleep}}$ are the time and power consumption during the \cgls{edrx} sleep period respectively, $T_{\text{PTW}}$ is the \cgls{ptw} length, $t_{\text{eDRX}}^{\text{cycle}}$ is the length of a \cgls{edrx} cycle, and $E_{\text{iDRX}}^{\text{cycle}}$ is the energy spent in an \cgls{idrx} cycle. 

\subsubsection{\cgls{psm}} \label{sec:psm}
\cgls{psm} is another power saving feature for \cgls{nbiot} and \cgls{ltem}. In \cgls{psm} the \cgls{ue} alternates between a deep sleep state and a period when the \cgls{ue} is reachable by the network. There are two main timers associated with \cgls{psm}: $T_{\text{3324}}$ and $T_{\text{3412}}$. The $T_{\text{3324}}$ timer determines the period when the \cgls{ue} is reachable, during which the \cgls{ue} can be operating either with \cgls{idrx} or \cgls{edrx}. The energy consumption model of \cgls{idrx} and \cgls{edrx} derived above can be reused within the period of $T_{\text{3324}}$. The \cgls{ue} will shut down all \cgls{as} functions and go into deep sleep upon the expiration of the $T_{\text{3324}}$ timer. The \cgls{ue} remains in deep sleep until the $T_{\text{3412}}$ timer expires, after which the \cgls{ue} will perform a \cgls{tau} procedure. It is observed from measurements that the \cgls{tau} procedure can be energy expensive. From an energy consumption point of view, it is better to configure the \cgls{tau} periodicity (i.e., $T_{\text{3412}}$) long enough to reduce the number of \cgls{tau} occurrences, while still meeting the service requirements of its use case.

The energy consumption of \cgls{psm} with \cgls{idrx} can be calculated as:
\begin{align}
    E_{\text{PSM}}^{\text{cycle}} &= \ceil[\bigg]{\frac{T_{\text{3324}}}{t_{\text{iDRX}}^{\text{cycle}}}} \cdot E_{\text{iDRX}}^{\text{cycle}} + t_{\text{PSM}}^{\text{sleep}}\cdot P_{\text{PSM}}^{\text{sleep}} \label{eq:psm_energy} \\
    t_{\text{PSM}}^{\text{sleep}} &= T_{\text{3412}} - T_{\text{3324}} 
\end{align}
where $P_{\text{PSM}}^{\text{sleep}}$ is the power consumption during sleep period. When using \cgls{edrx} the terms $E_{\text{iDRX}}^{\text{cycle}}$ and $t_{\text{iDRX}}^{\text{cycle}}$ should be replaced with the \cgls{edrx} equivalent. The \cgls{tau} procedure is not included here, but will be modelled separately in Section~\ref{sec:procedureModel}.

It is observed from the measurements that the power consumption during sleep period in \cgls{psm} is lower than \cgls{drx} or \cgls{edrx}, as the device turns off the radio. The disadvantage is that the device has to reconnect to the network when wakes up, which is energy expensive. Therefore for \cgls{iot} applications not requiring frequent transmit, \cgls{psm} is more energy efficient.

\subsection{Key Parameters of States}
From the previous subsections, it is noticed that the calculation of the energy consumption in each \cgls{ue} state for \cgls{ltem} and \cgls{nbiot} is similar to each other, as both technologies are developed based on \cgls{lte} and share a lot of similarities. The main differences in the calculation of energy consumption are in the \cgls{tx} and \cgls{rx} states. Specifically, for \cgls{tx} state the uplink resources in \cgls{ltem} are allocated in terms of \cgls{prb}s with 1 ms time resolution (subframe length). While in \cgls{nbiot} the uplink resources are allocated in terms of \cgls{ru}s with different lengths depending on the transmission option. For \cgls{rx} state, the downlink resources are allocated in terms of \cgls{prb}s with 1 ms time resolution for both \cgls{ltem} and \cgls{nbiot}. But \cgls{nbiot} only supports 1 \cgls{prb} resource allocation in downlink, while the default resource allocation for \cgls{ltem} is 6 \cgls{prb}s in downlink. Besides, different number of repetitions are supported in \cgls{ltem} and \cgls{nbiot}. These differences in the resource allocation and the number of repetitions effectively affect the required \cgls{tx} and \cgls{rx} time, resulting in different data rates and energy consumption between the two technologies. The key parameters for the calculation of the energy consumption in each \cgls{ue} state are summarized in \autoref{tab:StatesKeyParameters}, including physical layer transmission parameters and network configuration parameters.

\begin{table}[ht]
    \centering
    \begin{tabular}{|l|l|}
    \hline
    \textbf{UE States} & \textbf{Key Parameters}    \\ 
    \hline
    \textbf{TX} & Payload size, TX power, MCS, Repetitions, \\
    & allocated RUs (NB-IoT), allocated PRBs (LTE-M), \\
    & RU length (NB-IoT), Subframe length (LTE-M) \\
    \hline
    \textbf{RX} & Payload size, MCS, allocated PRBs, Repetitions,  \\
    & Subframe length  \\
    \hline
    \textbf{iDRX} &  iDRX Cycle, OnDurationTimer, Paging Repetitions  \\
    \hline
    \textbf{cDRX} & cDRX Cycle, OnDurationTimer \\
    \hline
    \textbf{eDRX} & eDRX Cycle, Paging Time Window  \\
    \hline
    \textbf{PSM} & $T_{\text{3324}}$, $T_{\text{3412}}$ \\
    \hline
    \end{tabular}
    \caption{Key parameters for the calculation of energy consumption in each \cgls{ue} state} 
    \label{tab:StatesKeyParameters}
\end{table}

It is of great importance to properly configure these parameters not only to ensure a good link performance such as radio coverage and throughput, but also to save \cgls{ue} energy consumption while satisfying the service requirements. The general guideline on how to configure the physical transmission parameters in the \cgls{tx} state, which is the most energy consuming state, is detailed in Section~\ref{sec:coverage}. The analysis of some of the important network configuration parameters which impact the \cgls{ue}'s batter lifetime effectively, such as \cgls{tau} periodicity ($T_{\text{3412}}$ timer), is discussed in Section~\ref{sec:results}.

\section{Energy Consumption of UE Procedures} \label{sec:procedureModel}
The transition among the \cgls{ue} states is dictated by higher layer protocol procedures. The main procedures in \cgls{nbiot} and \cgls{ltem} are similar to each other and have been introduced in Section~\ref{subsec:procedures}. Each procedure can be decomposed into a sequence of uplink and downlink message exchanges. By combining the associated TX (uplink transmission) and RX (downlink reception) state model together with the corresponding message size, the energy consumption in each procedure can be calculated.

\subsection{Message Exchange in Different Procedures}
The synchronization procedure is not explicitly modeled in this paper because it varies from device to device and is heavily dependent on the implementation. Instead, the energy and time consumption for the synchronization of a specific device is obtained from measurement. The measurement starts from the beginning of synchronization and ends before the occurrence of the first \cgls{prach}.

The measured message exchanges and the corresponding message size for the Attach, Service Request, \cgls{rrc} Release/Resume, and \cgls{tau} procedures are listed in \autoref{tab:nasattach,tab:SR,tab:rrcrelease,tab:tau}, recorded from a specific \cgls{nbiot} device N211 and an \cgls{ltem} device R410M. These two devices are the ones considered for measurements and the validation. A general description of these two devices are given in Section~\ref{sec:results}. 

The measured message sequence is in accordance with the protocol specifications defined in \cite{TS36331}. From the measurements it is observed that the size of the message might be device and network dependent, as the implementation of the protocol is device specific and some messages are dependent on the configuration of the network. It is also worth mentioning that the user data can be transmitted within the services request by using the \cgls{ciot} \cgls{eps} Optimization, thereby reducing the signalling overhead by skipping the \cgls{eps} bearers establishment~\cite{nbiot}. 

\begin{table}[tb]
    \centering
    \begin{tabularx}{\linewidth}{|X|p{1.4cm}|p{1.4cm}|p{1.2cm}|}
    	\hline
    	\textbf{Messages} & \multicolumn{1}{l|}{\textbf{NB-IoT}} & \multicolumn{1}{l|}{\textbf{LTE-M}} & \multicolumn{1}{l|}{\textbf{Direction}} \\
    	\hline
    	Random Access Preamble & \multicolumn{1}{l|}{RAP} & \multicolumn{1}{l|}{RAP} & \multicolumn{1}{l|}{uplink} \\
    	\hline
    	Random Access Response   & \multicolumn{1}{l|}{\SI{104}{\bit}} & \multicolumn{1}{l|}{\SI{56}{\bit}} & \multicolumn{1}{l|}{downlink} \\
    	\hline
    	RRC Connection Request & \multicolumn{1}{l|}{\SI{88}{\bit}} & \multicolumn{1}{l|}{\SI{72}{\bit}} & \multicolumn{1}{l|}{uplink} \\
    	\hline
    	RRC Connection Setup & \multicolumn{1}{l|}{\SI{144}{\bit}} & \multicolumn{1}{l|}{\SI{336}{\bit}} & \multicolumn{1}{l|}{downlink} \\
    	\hline
    	RRC Connection Complete & \multicolumn{1}{l|}{\SI{424}{\bit}} & \multicolumn{1}{l|}{\SI{656}{\bit}} & \multicolumn{1}{l|}{uplink} \\
    	\hline
    	Attach Request & \multicolumn{1}{l|}{\SI{256}{\bit}} & \multicolumn{1}{l|}{\SI{768}{\bit}} & \multicolumn{1}{l|}{uplink} \\
    	\hline
    	Identity Request & \multicolumn{1}{l|}{\SI{96}{\bit}} & \multicolumn{1}{l|}{\SI{40}{\bit}} & \multicolumn{1}{l|}{downlink} \\
    	\hline
    	Identity Response & \multicolumn{1}{l|}{\SI{176}{\bit}} & \multicolumn{1}{l|}{\SI{128}{\bit}} & \multicolumn{1}{l|}{uplink} \\
    	\hline
    	Authentication Request & \multicolumn{1}{l|}{\SI{432}{\bit}} & \multicolumn{1}{l|}{\SI{312}{\bit}} & \multicolumn{1}{l|}{downlink} \\
    	\hline
    	Authentication Response & \multicolumn{1}{l|}{\SI{264}{\bit}} & \multicolumn{1}{l|}{\SI{112}{\bit}} & \multicolumn{1}{l|}{uplink} \\
    	\hline
    	Security Command & \multicolumn{1}{l|}{\SI{328}{\bit}} & \multicolumn{1}{l|}{\SI{232}{\bit}} & \multicolumn{1}{l|}{downlink} \\
    	\hline
    	Security Complete & \multicolumn{1}{l|}{\SI{240}{\bit}} & \multicolumn{1}{l|}{\SI{176}{\bit}} & \multicolumn{1}{l|}{uplink} \\
    	\hline
    	Attach Accepted & \multicolumn{1}{l|}{\multirow{2}{*}{\SI{1080}{\bit}}} & \multicolumn{1}{l|}{\SI{792}{\bit}} & \multicolumn{1}{l|}{downlink} \\
    	\cline{1-1}\cline{3-4}UE Enquiry &       & \multicolumn{1}{l|}{\SI{208}{\bit}} & \multicolumn{1}{l|}{downlink} \\
    	\hline
    	UE Cabability & \multicolumn{1}{l|}{\SI{128}{\bit}} & \multicolumn{1}{l|}{\SI{256}{\bit}} & \multicolumn{1}{l|}{uplink} \\
    	\hline
    	Attach Complete & \multicolumn{1}{l|}{\SI{240}{\bit}} & \multicolumn{1}{l|}{\SI{256}{\bit}} & \multicolumn{1}{l|}{uplink} \\
    	\hline
    	EMM Info & \multicolumn{1}{l|}{\SI{488}{\bit}} & \multicolumn{1}{l|}{\SI{408}{\bit}} & \multicolumn{1}{l|}{downlink} \\
    	\Xhline{1pt}
    	Total Uplink &\SI{1816}{\bit}    & \SI{2424}{\bit}   &  \\
    	\hline
    	Total Downlink & \SI{2672}{\bit}   &  \SI{2384}{\bit}  &  \\
    	\hline
    \end{tabularx}
    \caption{Message exchange in the \textbf{Attach} procedure, measured from N211 (\cgls{nbiot}) and R410M (\cgls{ltem}).
    }
    \label{tab:nasattach}
\end{table}

\begin{table}[tb]
    \centering
    \begin{tabularx}{\linewidth}{|X|p{1.4cm}|p{1.4cm}|p{1.2cm}|}
		\hline
		\textbf{Messages}       & \textbf{\cgls{nbiot}} & \textbf{\cgls{ltem}} & \textbf{Direction} \\ \hline
		Random Access Preamble & RAP &RAP & uplink \\ \hline
		Random Access Response &\SI{104}{\bit}                  &\SI{56}{\bit}    & downlink                    \\ \hline
		\cgls{rrcReq}           &\SI{88}{\bit}                   &\SI{72}{\bit}     & uplink                    \\ \hline
		\cgls{rrcSet}           &\SI{144}{\bit}                  &\SI{336}{\bit}      & downlink                    \\ \hline
		\cgls{rrcComp}  &\SI{424}{\bit}                       & \SI{656}{\bit}    & uplink       \\ \hline
		Service request + UL data &\SI{56}{\bit} + data                       & N/A    & uplink       \\ \hline
		Service Accept   & \SI{176}{\bit}                             & N/A        & downlink                    \\ \hline
		Service Accept ACK   & \SI{32}{\bit}                             & N/A       & uplink                    \\ \hline
		RRC Reconfig            & \SI{72}{\bit}                             & N/A       & downlink                    \\ \hline
		RRC Reconfig Complete   & \SI{16}{\bit}                             & N/A       & uplink            \\ \Xhline{1pt}	
		Total \glsentrylong{ul} & \SI{616}{\bit}  + data        & \SI{728}{\bit}       & \\ \hline
		Total \glsentrylong{dl} & \SI{496}{\bit}                  &  \SI{392}{\bit}     & \\ \hline
	\end{tabularx}
    \caption{Message exchange in the \textbf{Service Request} procedure, measured from N211 (\cgls{nbiot}) and R410M (\cgls{ltem}).
    }
    \label{tab:SR}
\end{table}

\begin{table}[t]
    \centering
    \begin{tabularx}{\linewidth}{|X|p{1.4cm}|p{1.4cm}|p{1.2cm}|}
		\hline
		\textbf{Messages}       & \textbf{\glsentryshort{nbiot}} & \textbf{\glsentryshort{ltem}} & \textbf{Direction} \\ \hline
		\cgls{rrc} Release              & \SI{72}{\bit}                 & \SI{96}{\bit}                 & downlink                   \\ \hline
		\cgls{rrc} Resume              & \SI{72}{\bit}                 & \SI{96}{\bit}                 & downlink                   \\ \hline
		\cgls{ack}              & \SI{32}{\bit}                  &    \SI{32}{\bit}                           & uplink                    \\ \hline
	\end{tabularx}
    \caption{Message exchange in the \textbf{RRC Release/Resume} procedure, measured from N211 (\cgls{nbiot}) and R410M (\cgls{ltem}).}
    \label{tab:rrcrelease}
\end{table}

\begin{table}[t]
    \centering
    \begin{tabularx}{\linewidth}{|X|p{1.4cm}|p{1.4cm}|p{1.2cm}|}
		\hline
		\textbf{Messages}       & \textbf{\glsentryshort{nbiot}} & \textbf{\glsentryshort{ltem}} & \textbf{Direction} \\ \hline
		Random Access Preamble & RAP &RAP & uplink \\ \hline
		Random Access Response &\SI{104}{\bit}                  &\SI{56}{\bit}    & downlink                    \\ \hline
		\cgls{rrcReq}           &\SI{88}{\bit}                   &\SI{72}{\bit}     & uplink                    \\ \hline
		\cgls{rrcSet}           &\SI{144}{\bit}                  &\SI{336}{\bit}      & downlink                    \\ \hline
		\cgls{rrcComp}  &\SI{424}{\bit}                       & \SI{656}{\bit}    & uplink       \\ \hline
		\cgls{tau} Request  &\SI{144}{\bit}                       & \SI{224}{\bit}    & uplink       \\ \hline
		\cgls{tau} Accept   &\SI{448}{\bit}                  &\SI{512}{\bit}      & downlink         \\ \hline
		\cgls{tau} Complete  &\SI{80}{\bit}                       & \SI{112}{\bit}    & uplink       \\ \hline
		\cgls{rrc} Release   & \SI{72}{\bit}                 & \SI{96}{\bit}      & downlink      \\ \hline
		\cgls{ack}           & \SI{32}{\bit}                  & \SI{32}{\bit}    & uplink         \\ \Xhline{1pt}
		Total \glsentrylong{ul} & \SI{768}{\bit}              &  \SI{1096}{\bit}  & \\ \hline
    	Total \glsentrylong{dl} & \SI{768}{\bit}              & \SI{1000}{\bit}   & \\ \hline
	\end{tabularx}
    \caption{Message exchange in the \textbf{TAU} procedure, measured from N211 (\cgls{nbiot}) and R410M (\cgls{ltem}).}
    \label{tab:tau}
\end{table}

There are delays during the communication between the eNB and the \cgls{ue} for postprocessing of the received information and preparation for the next transmission. These delays should also be taken into consideration in the calculation of the energy consumption in each procedure. \autoref{tab:DelayType} summarizes the measured delays between different message exchanges.

\begin{table}[tb]
	\centering
	\begin{tabular}{l|ll}
		\textbf{Delay Type}    & \textbf{N211} & \textbf{R410M} \\ \hline
% 		\textbf{\cgls{tx} $\rightarrow$ \glss{dci}}     & 3 & 1\\
        RAP (TX) $\rightarrow$ DCI (RX)    & 3 [\si{\milli\second}] & 1 [\si{\milli\second}] \\
        DCI (RX) $\rightarrow$ RAR (RX)     & 4 [\si{\milli\second}] & 1 [\si{\milli\second}]\\
        RAR (RX) $\rightarrow$ RRC Request (TX)     & 8 [\si{\milli\second}] & 1 [\si{\milli\second}]\\
        RRC Request (TX) $\rightarrow$ DCI (RX)     & 3 [\si{\milli\second}] & 1 [\si{\milli\second}]\\
        DCI (RX) $\rightarrow$ RRC Setup (RX)     & 4 [\si{\milli\second}] & 1 [\si{\milli\second}]\\
		RRC Setup (RX) $\rightarrow$ DCI (RX)     & 12 [\si{\milli\second}] & 1 [\si{\milli\second}]\\
		DCI (RX) $\rightarrow$ RRC Setup Complete (TX)     & 8 [\si{\milli\second}] & 1 [\si{\milli\second}]   \\
		RRC Setup Complete (TX) $\rightarrow$ DCI (RX) & 3 [\si{\milli\second}] & 1 [\si{\milli\second}] \\
		DCI (RX) $\rightarrow$ Service Accept (TX) & 4 [\si{\milli\second}] & 1 [\si{\milli\second}] \\
		Service Accept (TX) $\rightarrow$ DCI (RX) & 12 [\si{\milli\second}] & 1 [\si{\milli\second}]   \\ \hline
		DCI (RX) $\rightarrow$ Data (RX) & 4 [\si{\milli\second}] & 3 [\si{\milli\second}] \\ 
		DCI (RX) $\rightarrow$ Data (TX) & 8 [\si{\milli\second}] & 3 [\si{\milli\second}] \\ 
		Data (RX) $\rightarrow$ DCI (RX) & 12 [\si{\milli\second}] & 3 [\si{\milli\second}] \\
		Data (TX) $\rightarrow$ DCI (RX) & 3 [\si{\milli\second}] & 1 [\si{\milli\second}] \\
		DCI (RX) $\rightarrow$ S1-Release & 4 [\si{\milli\second}] & 1 [\si{\milli\second}] \\ \hline
% 		\textbf{Data (TX) $\rightarrow$ ACK (RX)} & 3 & 1 \\ 
% 		\textbf{Data (RX) $\rightarrow$ ACK (TX)} & 12 & 3 \\ \hline
		SR $\rightarrow$ SR & N/A & 40 [\si{\milli\second}] \\
		SR $\rightarrow$ DCI (RX) & N/A & 3 [\si{\milli\second}]
	\end{tabular}
	\caption{The delays between different message exchanges, measured from N211 (\cgls{nbiot}) and R410M (\cgls{ltem}).
	}
	\label{tab:DelayType}
\end{table}

\subsection{Energy Consumption in the Procedure}
For each procedure, the number of exchanged messages and the corresponding message size are required to calculate the energy consumption. That information can be obtained for example from measurements as listed in \autoref{tab:nasattach,tab:SR,tab:rrcrelease,tab:tau}. The energy consumption in each procedure can be calculated as the sum of the energy cost of the exchanged messages and the energy cost of the delays between the message exchange, given as:
\begin{align} \label{eq:Eprocedure}
	E_{\text{Proced}} &= \sum_{i=1}^{I} \bigg( E^i_\text{Msg}(d_i) + E^i_\text{DCI} \bigg) + P_\text{delay} \cdot \sum_{i=1}^{I-1} t^i_\text{delay}
\end{align}
where $I$ is the total number of exchanged messages in the procedure excluding DCIs, $E^i_\text{Msg}(d_i)$ is the energy cost of transmitting (uplink) or receiving (downlink) message $i$ with payload size $d_i$ in bits, $E^i_\text{DCI}$ the energy cost of receiving the $i$th \cgls{dci}, $P_\text{delay}$ is the power consumption during the delay, and $t^i_\text{delay}$ is the delay between message $i$ and $i+1$.

$E^i_\text{Msg}(d_i)$ can be calculated based on the message type (e.g., uplink or downlink) by using the corresponding TX and RX state model described in Section~\ref{sec:stateModel}.  $t^i_\text{delay}$ can be obtained from \autoref{tab:DelayType}. From the measurements it is observed that $P_\text{delay}$ is device and technology dependent. Specifically, it is equivalent to the \gls{rx} state in R410M (\cgls{ltem}), while it is equivalent to the \cgls{cdrx} idle state in N211 (\cgls{nbiot}).

In \cgls{nbiot} the energy consumption of signalling overhead is negligible as it is much smaller compared to data transmissions. In \cgls{ltem}, the energy consumption in the waiting time between the signalling messages can be costly, for example the delay between the two Scheduling Request (SR) in \cgls{ltem} is observed with \SI{40}{\milli \second} interval.

%The time consumption in each procedure can be calculated in a similar way:
%\begin{align}
%	t_{\text{Procedure}} &= \sum_{i=1}^{I} %t^i_\text{Msg}(d_i) + \sum_{i=1}^{I-1} %t^i_\text{DRX\&DCI}
%\end{align}

%In \cgls{nbiot} the energy consumption of signalling overhead can be disregarded because it is much smaller as compared to data transmission. However, this is not the case for \cgls{ltem}. Furthermore, the energy consumption in the waiting time between the transmission and reception of signalling messages is higher in \cgls{ltem}, which is energy costly for \cgls{ltem} operating in \gls{hdfdd} where the period between \gls{pucch} resources is observed with \SI{40}{\milli \second} interval.

%% file: 03_Scenario.tex
\section{Battery Lifetime Estimation}\label{sec:estimation}
The calculation of the energy consumption of each \cgls{ue} state and procedure is derived in Section~\ref{sec:stateModel} and \ref{sec:procedureModel}, respectively. By combining different components of \cgls{ue} states and procedures, the energy consumption of any \cgls{ue} behaviour can be calculated. The other two prerequisites required to estimate the battery lifetime of an \cgls{iot} device are the coverage scenario which determines the physical transmission parameters, and the traffic profile which defines the traffic characteristics of an \cgls{iot} application. 

\subsection{Coverage Scenario} \label{sec:coverage}
The actual calculation of the proposed power consumption model is tightly associated with the configuration of physical layer transmission parameters such as \cgls{mcs}, the number of allocated \cgls{ru}s/\cgls{prb}s, and the number of repetitions, which further depends on the coverage level of the \cgls{ue}. 

To allow the \cgls{ue} to be served in different coverage conditions characterized by different path loss, \cgls{3gpp} Release 13 has defined three \cgls{ce} levels: \cgls{ce} level 0 (normal coverage with \cgls{mcl}$\approx$144 dB), \cgls{ce} level 1 (robust coverage with \cgls{mcl}$\approx$154 dB), and \cgls{ce} level 2 (extreme coverage with \cgls{mcl}$\approx$164 dB). The \cgls{ce} level is selected based on the channel conditions, and determines the physical layer transmission parameters.

In our measurement, three coverage scenarios, namely "Good", "Bad", and "Extreme", have been defined corresponding to a coupling loss of 140 dB, 150 dB, and 160 dB, respectively. Those values are derived based on the link budget calculations with configured uplink transmission parameters. Specifically, the allocated bandwidth in uplink is restricted to a single \cgls{prb} for \cgls{ltem} and a single subcarrier of 15 kHz for \cgls{nbiot}. The receiver noise figure is assumed to be 5 dB and the thermal noise density is $174$ \SI{}{dBm/Hz}. The \cgls{ue}'s maximum transmission power is 23 dBm. Based on these assumptions, the estimated \cgls{snr} without using repetitions for the good, bad, and extreme coverage scenarios are 10.24 dB, 0.24 dB and -9.76 dB for \cgls{nbiot}, and -0.55 dB, -10.55 dB and -20.55 dB for \cgls{ltem} \cite{coverageAnalysisLTEM}. When blind repetition with chase combining is used in \cgls{harq}, the effective received \cgls{snr} in linear scale after combining assuming infinite \cgls{tbs} can be calculated as \cite{LTESim}: 
\begin{equation}\label{eq:LinkAbstraction}
    \text{SNR}_{\text{combined}}=N \cdot\text{SNR}_{\text{RX}}
\end{equation}
where $N$ is the number of repetitions. The  number of repetitions in uplink can reach up to 128 for \cgls{nbiot}. In our analysis, the maximum number of repetitions is limited to 32 for realistic scenarios.

Once the received \cgls{snr} is calculated for each coverage scenario, the physical layer transmission parameters such as \cgls{mcs} is selected for a target \cgls{bler}. For \cgls{lte} systems, the target \cgls{bler} is set to be 10\% for data channels. The mapping between \cgls{snr} and \cgls{mcs} for a target \cgls{bler} can be found either from link level performance curves or from analytical approximations. 

Based on the link budget analysis mentioned above, the considered uplink physical layer transmission configurations for the three coverage scenarios for \cgls{nbiot} and \cgls{ltem} are listed in \autoref{tab:Configuration}. In order to make a fair comparison between \cgls{nbiot} and \cgls{ltem}, the physical layer transmission parameters are configured for each technology targeted for a similar coupling loss. The configuration for \cgls{ltem} in the extreme scenario is left blank as it can not reach the area with 160 dB coupling loss.

\begin{table}[h]
    \begin{tabular}{l|ccc|ccc}
         & \multicolumn{3}{c|}{\textbf{\cgls{nbiot}}} & \multicolumn{3}{c}{\textbf{\cgls{ltem}}} \\
        \textbf{Parameters} & \textbf{Good}  & \textbf{Bad}   & \textbf{Extreme} & \textbf{Good}  & \textbf{Bad}   & \textbf{Extreme}\\ \hline
        \textbf{\cgls{mcl}} [dB]             &  140 & 150 & 160 & 140 & 150 & 160 \\ \hline
       	%\textbf{Noise Power}      & -129,24     & -129,24      & -129,24       & -118.45      & -118.45       & -118.45\\ \hline
        \textbf{\cgls{mcs}} & 10 & 2 & 0 & 5 & 0 & - \\
        \textbf{Repetitions} & 1 & 8 & 32 & 2 & 16 & -\\ 
    \end{tabular}
    \caption{Considered uplink transmission configurations for \cgls{nbiot} and \cgls{ltem} in different coverage scenarios.}
    \label{tab:Configuration}
\end{table}

\subsection{Traffic Profile}
The traffic profile defines how often the \cgls{ue} transmits/receives data and how big the transmitted/received data is. It has a big impact on the battery lifetime of a device. The \cgls{iot} use cases and traffic patterns associated with some of the verticals have been defined in \cite{Mocnej}, which indicates that most of the \cgls{iot} traffic is uplink dominated with periodic traffic pattern. Therefore in this study, a deterministic uplink traffic model is used, assuming that the \cgls{ue} transmits $B$ \SI{}{Bytes} of uplink payload towards the eNB periodically at an average rate of $\lambda$ transmissions per hour.

\begin{figure}[tb]
	\centering
%	\begin{subfigure}{1\linewidth}
	\includegraphics[width=1\linewidth]{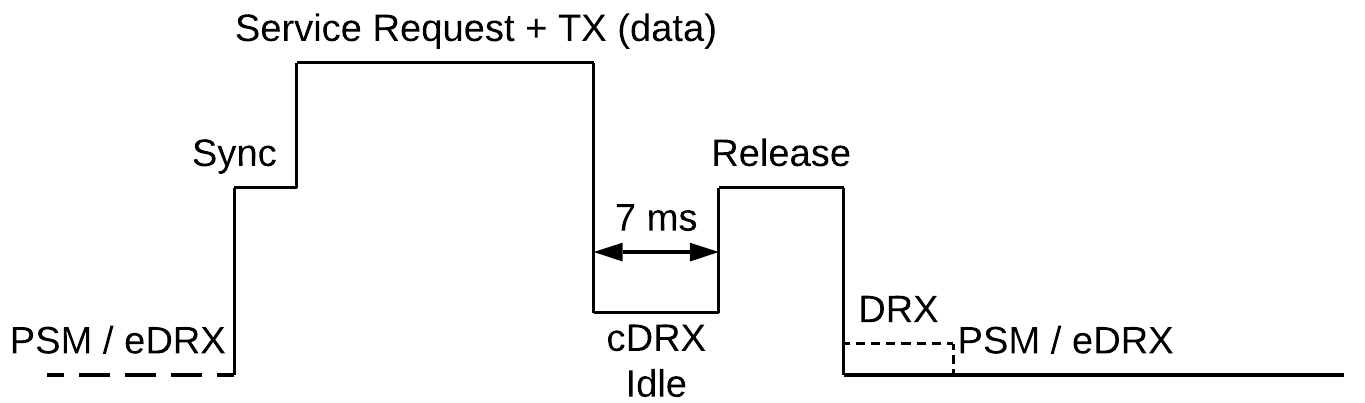}
%	\caption{Cycle A.}
%	\label{fig:batteryLife_cycleA}
%	\end{subfigure}
	\\ 
%	\begin{subfigure}{1\linewidth}
%	\includegraphics[width=1\linewidth]{Figures/BatteryLifeModeling/cycleB.pdf}
%	\caption{Cycle B.}
%	\label{fig:batteryLife_cycleB}
%	\end{subfigure}
	\caption{Illustration of a transmit cycle for \cgls{ciot}.}
	\label{fig:batteryLife_cycle}
\end{figure}

A transmit cycle is defined as the time interval between the start of a data transmission to the time instance right before the start of the next data transmission. The \cgls{ue} will go through certain procedures in a transmit cycle to establish and release a connection. An example of uplink \cgls{ue} transmit cycle that starts and ends in a power saving state is illustrated in \autoref{fig:batteryLife_cycle}. Upon waking up from \cgls{psm}, the \cgls{ue} first synchronizes to the network, then performs a service request together with uplink data transmission. After that the \cgls{ue} stays in the \cgls{cdrx} state for certain time until the release procedure is initiated by the network. Then the \cgls{ue} enters the power saving mode until the next uplink transmission is initiated or paging indicates a downlink transmission. In \autoref{fig:batteryLife_cycle} it is assumed that the \cgls{ue} is in \cgls{emmr} state. If the \cgls{ue} is in \cgls{emmd} state, e.g., the \cgls{ue} is powered up for the first time, the attach procedure is initiated instead of the service request procedure. 

\subsection{Battery Lifetime Modelling}
Once the energy consumption for each \cgls{ue} state and procedure has been determined, the physical layer transmission parameters have been set according to certain coverage scenario, and the traffic profile has been defined, the energy consumption of an \cgls{iot} device can be calculated and the corresponding battery lifetime can be estimated.

The energy consumption of the modem during a transmit cycle depicted in \autoref{fig:batteryLife_cycle} with a payload size of $B$ Bytes can be calculated as:  
\begin{align}
E_\text{cycle}(B) = E_\text{Sync} + E_{\text{SR}}(B) +  E_{\text{cDRX}} + E_{\text{Release}} +  E_{\text{PSM/eDRX}}^{\text{cycle}}
\end{align}
where $E_\text{Sync}$, $E_\text{SR}$, and $E_\text{Release}$ are the energy consumption in the synchronization, service request, and \cgls{rrc} release procedures respectively, which can be obtained using \autoref{eq:Eprocedure}. $E_{\text{cDRX}}$ and $E_{\text{PSM/eDRX}}^{\text{cycle}}$ are the energy consumption when the \cgls{ue} is in \cgls{cdrx} state and \cgls{idrx} state using either \cgls{psm} or \cgls{edrx} respectively, which can be obtained following the energy consumption approach described in Section~\ref{sec:state_ecmc} and Section~\ref{sec:state_ecmi}.

Let $\lambda$ denote the average transmission rate per hour. The average energy consumption of the modem during one hour can be calculated as:
\begin{align}
E_\text{hour}(\lambda,B) = E_\text{cycle}(B) \cdot \lambda
\end{align}

The estimated battery lifetime (in hours) of an \cgls{iot} device can be calculated as:
\begin{align}
    L(\lambda,B) = \dfrac{C_{\text{bat}}\cdot \text{SF}_{\text{bat}}}{E_\text{hour}(\lambda,B)+E_\text{device}}
    \label{eqn:lifetime}
\end{align}
where $C_{\text{bat}}$ is the battery capacity in [Wh], $\text{SF}_{\text{bat}}$ is the battery safety factor accounting for self-discharge effect, and $E_{\text{device}}$ is the sensor circuitry average energy consumption per hour, i.e. the energy consumption besides the modem.

%% file: 04_Results.tex
\section{Measurements and Validation}\label{sec:results}
A testbed has been developed to validate the proposed power consumption model as well as to estimate the battery lifetime of an \cgls{iot} device.

\subsection{Measurement Setup}
The power consumption of two commercial \cgls{dut}s have been measured, namely U-Blox EVK-N211 and U-Blox EVK-R410M. The N211 is an \cgls{nbiot} device, while R410M can be connected to both \cgls{nbiot} and \cgls{ltem} networks.

\begin{figure}[tb]
    \centering
    \includegraphics[width=\linewidth]{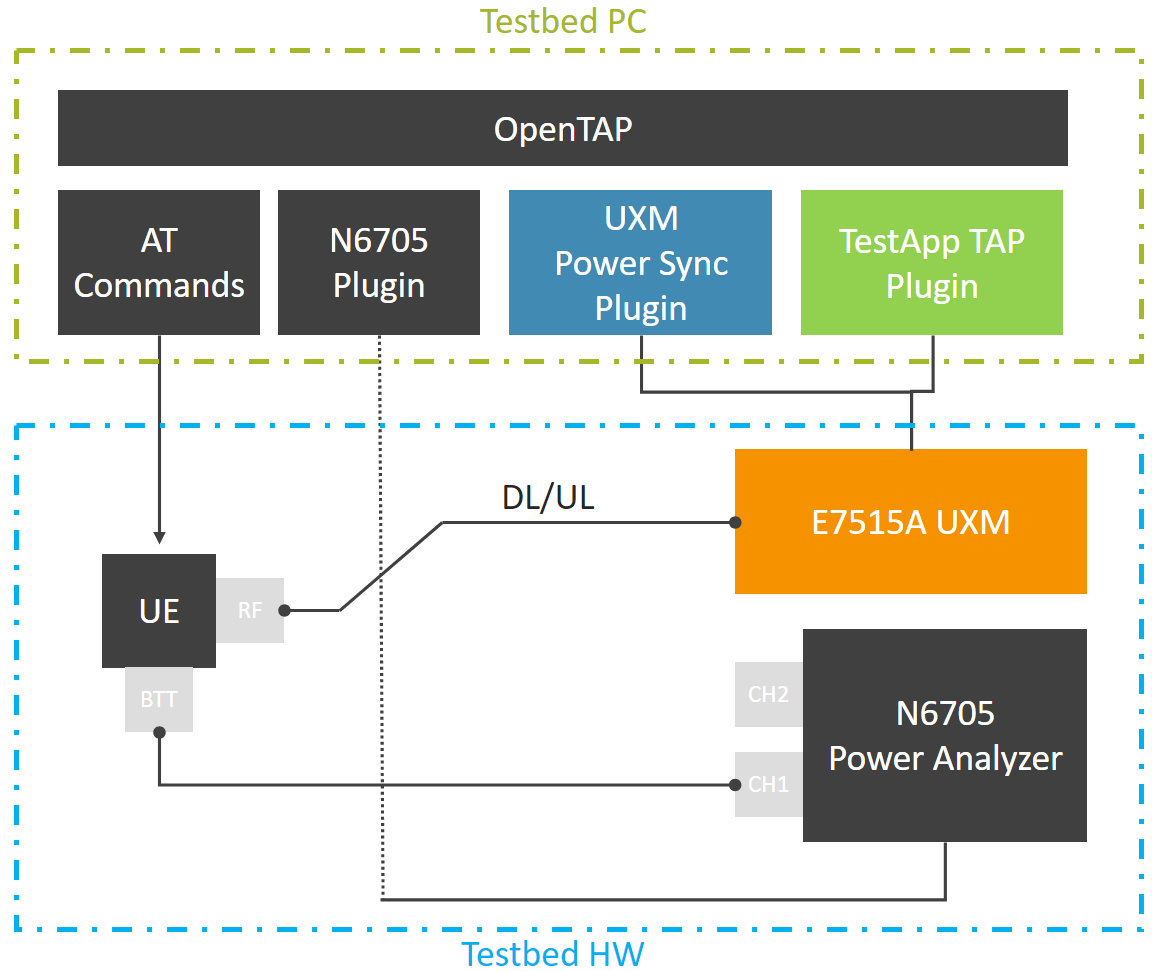}
    \caption{Testbed setup.}
    \label{fig:testbed_setup}
\end{figure}

The measurement setup is depicted in \autoref{fig:testbed_setup}. The \cgls{dut}'s antenna port is connected via cables to a Keysight E7515A UXM Wireless Test Set, which is a standard-compliant base station emulator supporting both Release 14 \cgls{nbiot} and \cgls{ltem} features with debugging capabilities. The \cgls{dut} is also connected to a Keysight N6705B DC Power Analyzer which acts as both a power supply and a sensor for battery drain measurements. The measurement setup is controlled by Keysight's \cgls{tap}, which provides interfaces to both the measurement equipments and the \cgls{dut}, and orchestrates the behaviour of different components by using \cgls{tap} test plans. The measurement setup is capable of synchronizing the network logs and power consumption measurements with $\leq$ \SI{1}{\milli\second} accuracy. This allows for in-depth analysis of measurements to quantify the power consumption. 

\subsection{Characterization of the Modem}
The analytical energy consumption model for each \cgls{ue} state described in Section~\ref{sec:stateModel} depends on the inputs of the modem characterization information such as the \cgls{ue} power level in different states, e.g., $P_{\text{TX}}$, $P_{\text{TX}}^{\text{gaps}}$, $P_{\text{RX}}$, $P_{\text{RX}}^{\text{gaps}}$, $P_{\text{cDRX}}^{\text{sleep}}$, $P_{\text{PSM}}^{\text{sleep}}$, etc. Therefore it is important to accurately characterize the power consumption of the modem so that the derived energy consumption model can work correctly. A test case has been executed for each \cgls{ue} state. From each measurement, only the power or energy consumption related to the target \cgls{ue} state is extracted and averaged over certain period. The measured power or energy consumption of N211 and R410M in different states are summarized in \autoref{tab:PowerEnergyConsumption}, which serves as inputs to the equations in Section~\ref{sec:stateModel}. Typically this modem characterization information is available from the vendors. 

\begin{table}[tb]
\begin{tabular}{ll|c|cc}
\multirow{2}{*}{\textbf{State}} & \multirow{2}{*}{\textbf{Symbol}}  & \textbf{N211}   & \multicolumn{2}{c}{\textbf{R410M}}        \\
                                &                                                                                & \textbf{NB-IoT} & \textbf{NB-IoT} & \textbf{LTE-M} \\ \hline
\multirow{2}{*}{\textbf{TX} (23 \text{dBm})}    & $P_{\text{TX}}$ [\si{mW}]         & 742.858         & 1421.391        & 1322.157       \\
                                & $P_{\text{TX}}^{\text{gaps}}$ [\si{mW}]                   & 153.6           & 168.800         & N/A             \\ \hline
\multirow{2}{*}{\textbf{RX}}    & $P_{\text{RX}}$ [\si{mW}]   & 222.134         & 174.427         & 335.607        \\
                                & $P_{\text{RX}}^{\text{gaps}}$ [\si{mW}]              & 177.422         & 174.097         & N/A             \\ \hline
\multirow{3}{*}{\textbf{cDRX}}  & $P_{\text{cDRX}}^{\text{sleep}}$ [\si{mW}]  & 21.337          & 34.476          & 34.458         \\
                                & $E_{\text{onDur}}$ [\si{mJ}]    & 0.885           & 1.847           & 0.319          \\
                                & $t_{\text{onDur}}$ [\si{ms}]    & 7.926           & 9.518           & 0.998          \\ \hline
\multirow{3}{*}{\textbf{eDRX}}   & $P_{\text{eDRX}}^{\text{sleep}}$ [\si{mW}] & 0.0122          & 3.686          & 3.654         \\
                                & $\text{E}_{\text{onDur}}$ [\si{mJ}] & 0.326           & 0.180           & 0.241          \\
                                & $t_{\text{onDur}}$ [\si{ms}]    & 1.445           & 1.104           & 1.02           \\ \hline
\textbf{PSM}                    & $P_{\text{PSM}}^{\text{sleep}}$ [\si{\micro W}]     & 9.5             & 46              & 46             \\ \hline
\multirow{2}{*}{\textbf{Sync}}  & $E_{\text{sync}}$ [\si{mJ}] & 160             & 362             & 1095           \\
                                & $t_{\text{sync}}$ [\si{ms}] & 2200            & 1361            & 4740          
\end{tabular}
	\caption{The measured power consumption of N211 and R410M in different states.}
	\label{tab:PowerEnergyConsumption}
\end{table}

\begin{figure}[tb]
    \centering
    \includegraphics[width=\linewidth]{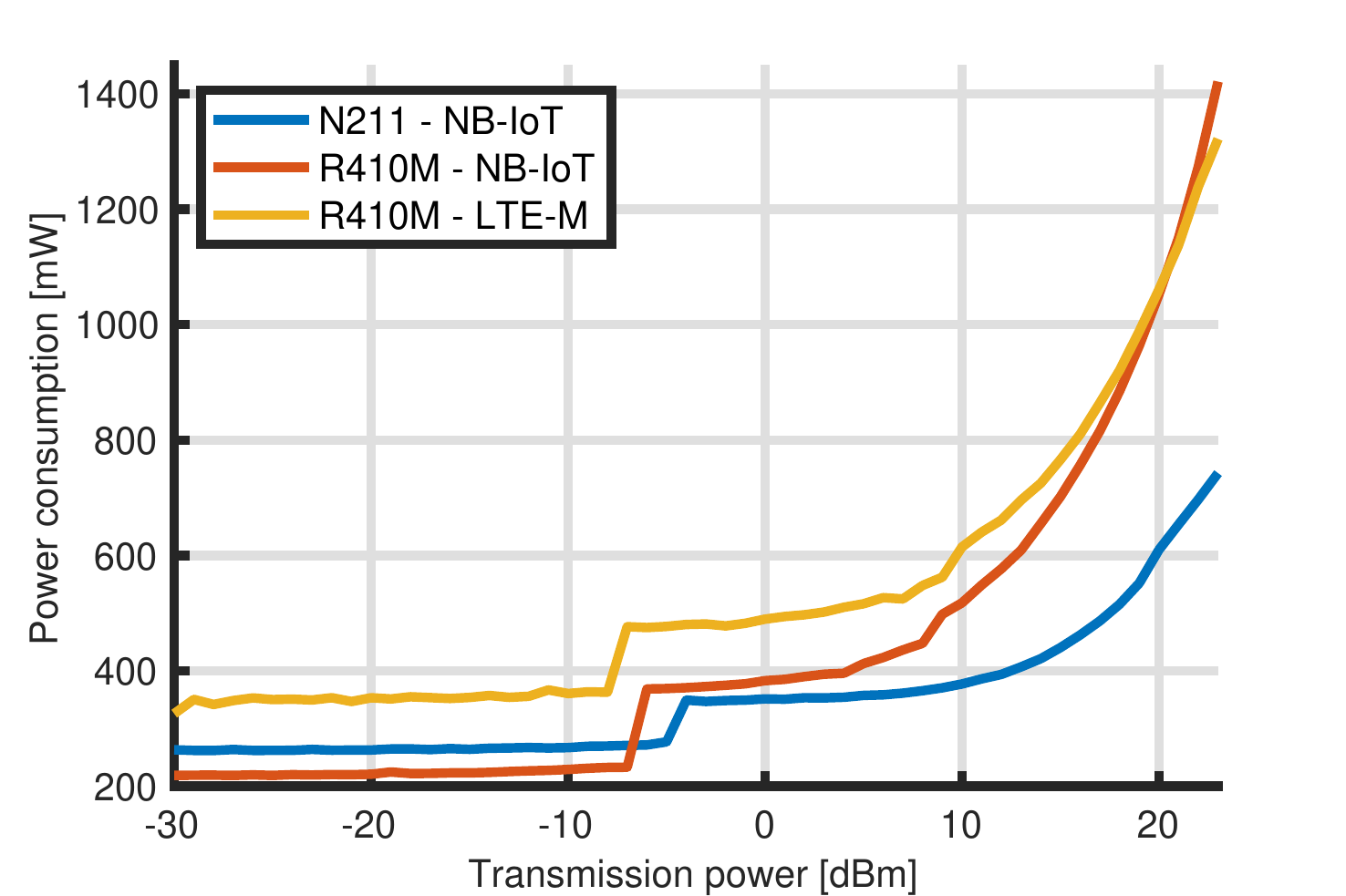}
    \caption{The measured power consumption of N211 and R410M in the TX state as a function of uplink transmission power.}
    \label{fig:txpower}
\end{figure}

It is observed from the measurements that the power or energy consumption of R410M is higher than N211 in most of the states, regardless of whether it is operating in \cgls{nbiot} or \cgls{ltem} mode, which is not surprising as the power consumption of a device is implementation specific. In particular, the measured power consumption of the two devices in the TX state as a function of uplink transmission power is plotted in \autoref{fig:txpower}. It can be seen that the power consumption curve can be split into two parts. The first part is when the power amplifier is not required, resulting in almost linear increase of the power consumption. The second part is when the power amplifier is used. In this case, the power consumption increases exponentially, which is due to the fact that the efficiency of the power amplifier decreases with the increase of the output power. Similar trend is observed for both devices, but the power consumption of R410M is higher than N211 when the uplink transmission power is larger than -8 dBm. This difference is much more visible when the uplink transmission power becomes higher.

\subsection{Model Validation}
To validate the accuracy of the proposed power consumption model, extensive measurement campaigns have been executed on both \cgls{nbiot} device N211 and  \cgls{ltem} device R410M with different configurations. Both short (with periodic transmit cycle of 1 hour) and long (with periodic transmit cycle of 24 hours) measurements have been performed with a fixed payload size of 100 Bytes. The $T_{\text{3324}}$ and $T_{\text{3412}}$ timers are used in \cgls{psm}, which is applied due to its relatively low power consumption during the sleep period as compared to \cgls{edrx}. The setting of the $T_{\text{3324}}$ timer has a tradeoff between low energy consumption and low response time to the application server. It is recommended by GSM Association (GSMA) that the $T_{\text{3324}}$ timer should best fit the \cgls{iot} use case, and the ratio between $T_{\text{3324}}$ Active Timer and $T_{\text{3412}}$ Extended Timer, calculated as $(T_{\text{3412}}-T_{\text{3324}})/T_{\text{3412}}$, should be $>90\%$ in order to achieve optimum battery savings \cite{gsma_ltem}. In our measurement,  the $T_{\text{3324}}$ timer is set to be 60 seconds, and the $T_{\text{3412}}$ timer for the \cgls{tau} periodicity is configured to be 2 hours which indicates that the \cgls{tau} procedure is not included in the short measurement, but is included in the long measurement. The battery capacity is set to be 5 Wh, assuming an ideal case without the self-discharge effect. Since the focus of this work is on the power consumption modelling of the modem, the power consumption of the sensor circuitry is not taken into account, which means that all of the available battery capacity is allocated to the modem. The available \cgls{mcs} index ranges from 0 to 10, and the number of repetitions for data channels ranges from 1 to 16, with different combinations between the two parameters. An example of the configured \cgls{mcs} index and repetition number for different scenarios are listed in \autoref{tab:Configuration}. All measurements are performed in LTE band 20 ($\sim 806$ MHz). For \cgls{nbiot}, a single tone with 15 kHz spacing is allocated in uplink and a single \cgls{prb} is allocated in downlink. For \cgls{ltem}, a single \cgls{prb} is allocated in uplink and 6 \cgls{prb}s are allocated in downlink. The number of allocated \cgls{ru}/\cgls{sf} is set to be 5 for both \cgls{nbiot} and \cgls{ltem}. The measurement settings used for model validation and battery lifetime estimations are summarized in \autoref{tab:assumptions}.

\begin{figure}[t]
    \centering
    \includegraphics[width=\linewidth]{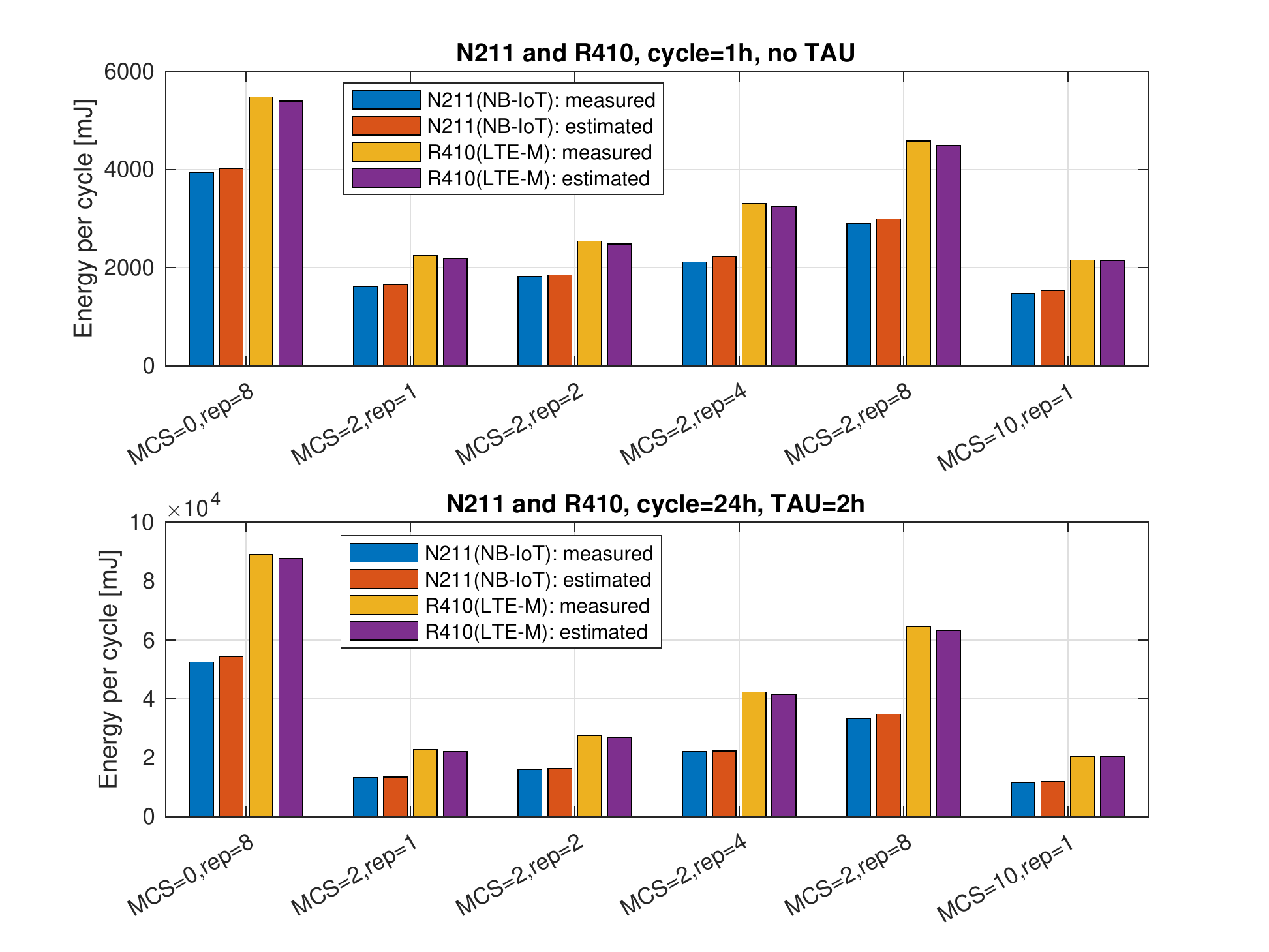}
    \caption{The measured and estimated energy consumption per transmit cycle (1 hour and 24 hours) for NB-IoT device N211 and LTE-M device R410M with different configurations.}
    \label{fig:validation}
\end{figure}

\begin{table}[h]
    \centering
	\begin{tabular}{l|c}
	 \textbf{Parameter} & \textbf{Setting} \\
	 \hline
	 \textbf{Carrier frequency} & LTE band 20 \\
     \textbf{UE transmit cycle} & 1 to 24 hour(s) \\
     \textbf{Payload size} & 100 Bytes \\
     \textbf{Battery capacity} & 5 Wh \\
     \textbf{Battery safety factor} & 1 \\
     \textbf{Uplink resource for NB-IoT} & single-tone with 15 kHz, 5 RUs \\
     \textbf{Downlink resource for NB-IoT} & 1 PRB, 5 SFs \\
     \textbf{Uplink resource for LTE-M} & 1 PRB, 5 SFs \\
     \textbf{Downlink resource for LTE-M} & 6 PRBs, 5 SFs \\
     \textbf{MCS Index for data channels} & \{0, 2, 5, 10\} \\
     \textbf{MCS Index for signalling messages} & \{0, 2\} \\
     \textbf{Repetitions for data channels} & \{1, 2, 4, 8, 16\} \\
     \textbf{Repetitions for signalling messages} & \{Repetitions for data\} x 2 \\
     \textbf{Power saving technique} & PSM \\
     \textbf{$T_{\text{3324}}$ timer} & 60 seconds \\
     \textbf{$T_{\text{3412}}$ timer (TAU periodicity)} & 2 hours \\
	\end{tabular}
	\caption{Measurement settings used for model validation and battery lifetime estimation.}
	\label{tab:assumptions}
\end{table}

\autoref{fig:validation} shows the measured and estimated energy consumption per transmit cycle (i.e. 1 hour and 24 hours) for both \cgls{nbiot} device N211 and  \cgls{ltem} device R410M with different configurations. If we compare the energy consumption between the two devices, it is shown clearly that the R410M consumes more energy as compared to N211, due to the reason that the energy consumption in each state is higher in R410M than in N211 as shown in \autoref{tab:PowerEnergyConsumption}. If we compare the energy consumption of the same device but with different configurations, it can be seen that the energy consumption decreases as the \cgls{mcs} index increases. Though not shown in the paper, it is found from the measurements that the average power consumption is independent of \cgls{mcs} for both uplink and downlink. However, for a fixed payload size and \cgls{tx} bandwidth, the selection of \cgls{mcs} affects the number of allocated \cgls{ru}s/subframes (i.e., transmission time), which means that for lower \cgls{mcs} the \cgls{ue} has to stay in the \cgls{tx} state for longer time, resulting in higher energy consumption. Also it is shown clearly that increasing the number of repetitions would increase the total energy consumption as expected. 

\autoref{fig:validation} clearly demonstrates that the proposed power consumption model matches very well with the measurement results for the selected two devices, regardless of different configurations of transit cycles and transmission parameters. The measured estimation error is within 5\% in all cases.

\subsection{Battery Lifetime Estimation}
Next we apply Eqn. (\ref{eqn:lifetime}) to estimate the battery lifetime with different configurations of traffic profiles, coverage scenarios (i.e., good, bad, and extreme defined in \autoref{tab:Configuration}), and network parameters. The same parameter settings listed in \autoref{tab:assumptions} are used for the battery lifetime estimation. 

\begin{figure}[t]
    \centering
    \includegraphics[width=\linewidth]{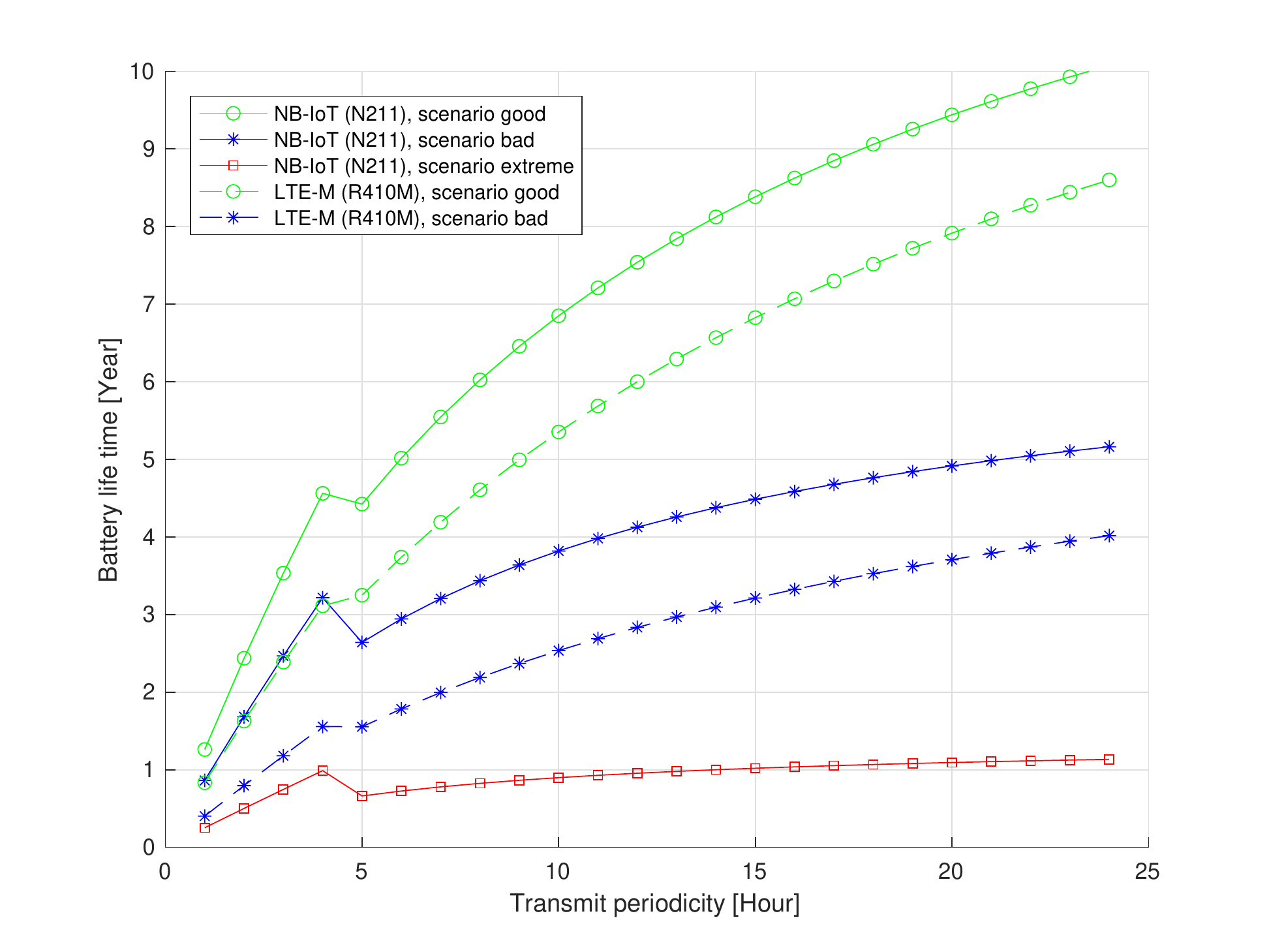}
    \caption{Estimated battery lifetime for NB-IoT device N211 with different payloads, coverage scenarios, and $T_{\text{3412}}$ timer, assuming the transmit cycle to be every 12 hours}
    \label{fig:lifetime_payload}
\end{figure}

\begin{figure}[t]
    \centering
    \includegraphics[width=\linewidth]{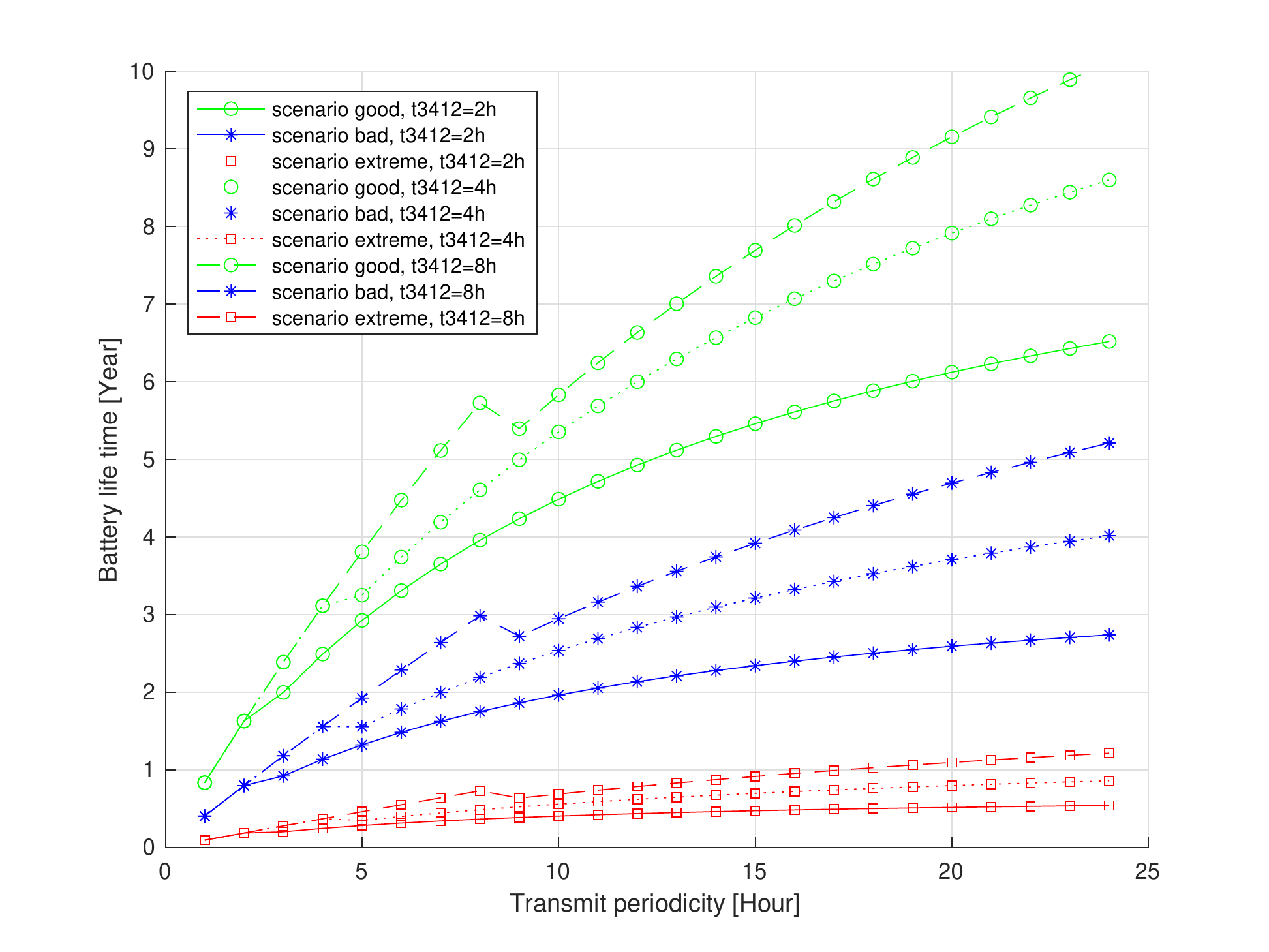}
    \caption{Estimated battery lifetime for NB-IoT device N211 with different transmit cycles, coverage scenarios, and $T_{\text{3412}}$ timer, assuming the payload size to be 100 Bytes}
    \label{fig:lifetime_cycle}
\end{figure}

\autoref{fig:lifetime_payload} and \autoref{fig:lifetime_cycle} show the estimated battery lifetime for NB-IoT device N211 versus different payload size and transmit cycles with different coverage scenarios and $T_{\text{3412}}$ timer settings, respectively. Reduce the payload size and/or increase the transmit cycle length would result in a longer battery lifetime in all scenarios as expected. Also it is clearly shown that the coverage scenario, which determines the PHY transmission parameters, has a great impact on the battery lifetime. Note that in \autoref{fig:lifetime_cycle}, there is a turning point when the transmit cycle equals the $T_{\text{3412}}$ timer. This is because when the transmit cycle is longer than the $T_{\text{3412}}$ timer (i.e. \cgls{tau} periodicity), the \cgls{tau} procedure will occur which consumes additional energy. When the transmit cycle is much longer than the TAU periodicity, the performing of the \cgls{tau} procedure can be energy expensive and will dominate the total energy consumption. That explains why in the extreme scenario in \autoref{fig:lifetime_cycle}, the battery lifetime does not increase very much with the increase of the transmit cycle. Increasing the \cgls{tau} periodicity will decrease the total energy consumption within one transmit cycle, as the occurrence of \cgls{tau} decreases, resulting in an increase of the battery lifetime as shown in \autoref{fig:lifetime_payload} and \autoref{fig:lifetime_cycle}.

\begin{figure}[t]
    \centering
    \includegraphics[width=\linewidth]{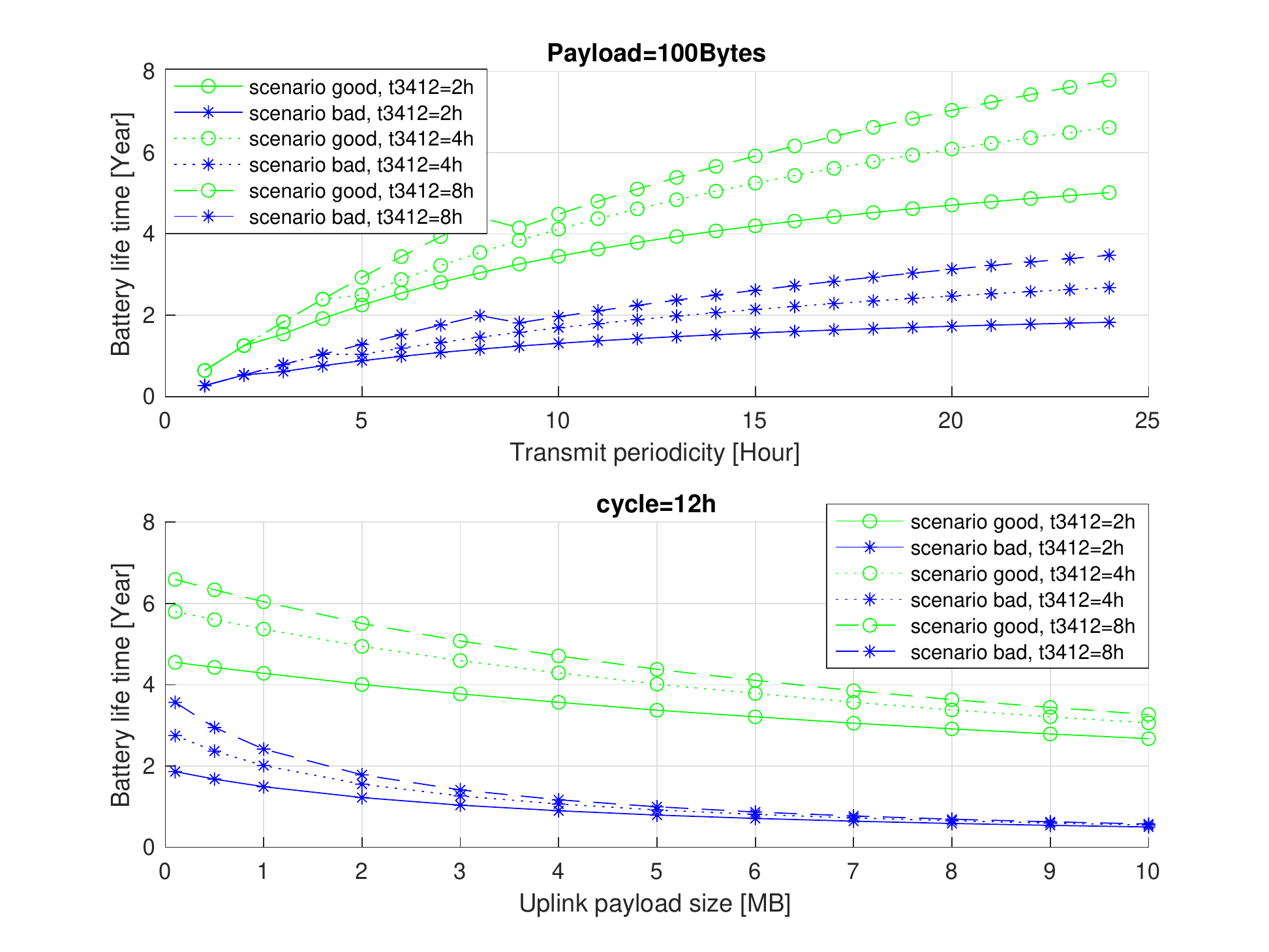}
    \caption{Estimated battery lifetime for LTE-M device R410M with different payloads, transmit cycles, coverage scenarios, and $T_{\text{3412}}$ timer}
    \label{fig:lifetime_ltem}
\end{figure}

The estimated battery lifetime for \cgls{ltem} device R410M versus different payloads and transmit cycles with different coverage scenarios and $T_{\text{3412}}$ timers is plotted in \autoref{fig:lifetime_ltem}. Similar behaviour has been observed in R410M as compared to N211.

\begin{figure}[t]
    \centering
    \includegraphics[width=\linewidth]{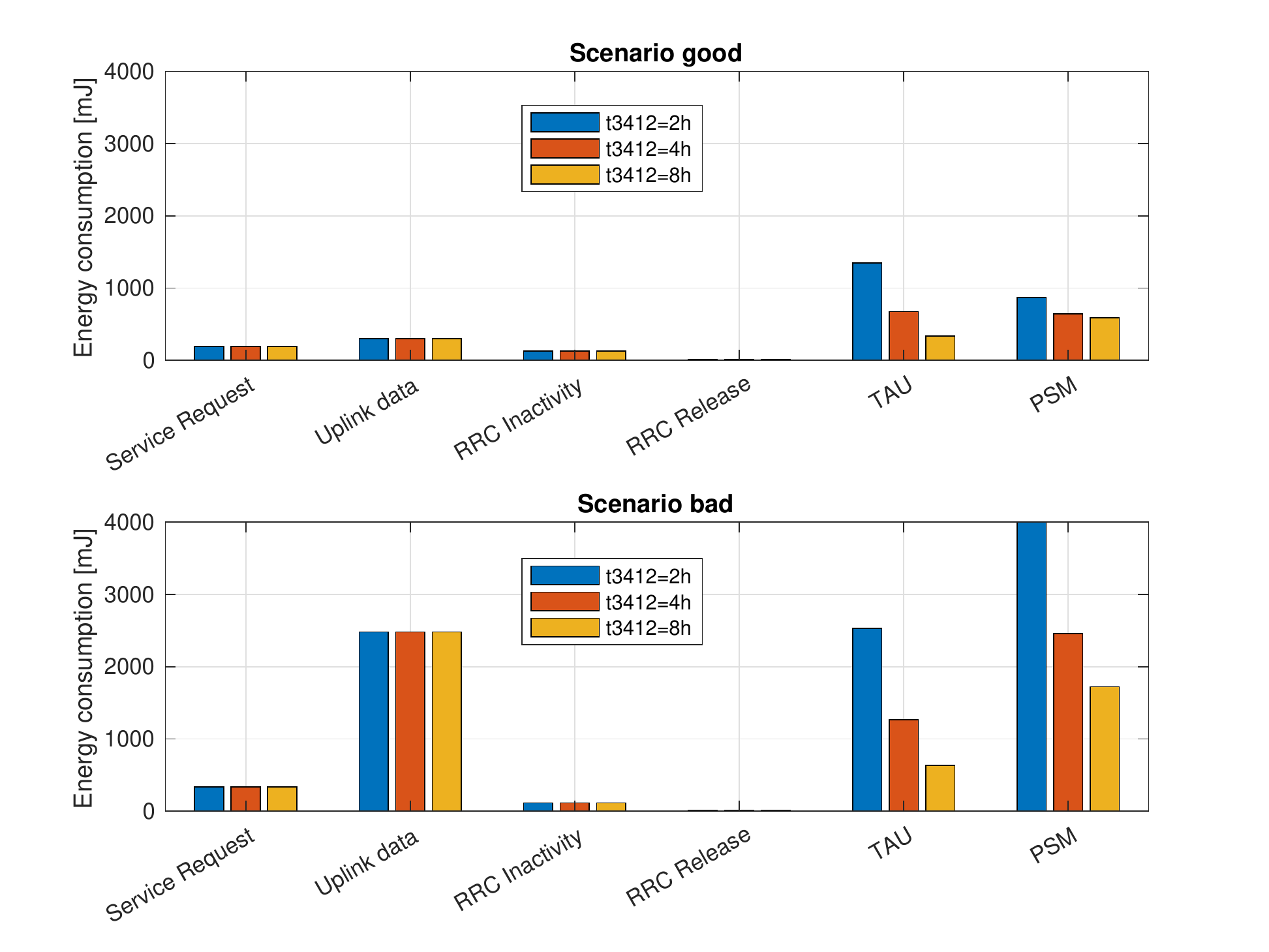}
    \caption{Energy consumption in different procedures and states during one transmit cycle (12 hours) for NB-IoT device N211, assuming the payload size to be 100 Bytes}
    \label{fig:energy_state}
\end{figure}

\autoref{fig:energy_state} shows the energy consumption in different procedures and states during one transmit cycle (12 hours) for NB-IoT device N211 under different scenarios and $T_{\text{3412}}$ timer settings, assuming the payload size to be 100 Bytes. Using lower \cgls{mcs} index and higher number of repetitions (e.g. the bad scenario) will cause higher energy consumption as compared to the good scenario, because of the increase in the uplink transmission time. It is shown in the figure that the $T_{\text{3412}}$ timer has no effect on the energy consumption on the service request, uplink data transmission, \cgls{rrc} inactivity and release procedures, but has big impact on the \cgls{tau} and \cgls{psm}. Increasing the  $T_{\text{3412}}$ timer would not only decrease the energy consumption in the \cgls{tau} procedure, but also in the \cgls{psm} procedure, because the $T_{\text{3412}}$ timer determines the period when the device is in sleep mode.

\begin{figure}[t]
    \centering
    \includegraphics[width=\linewidth]{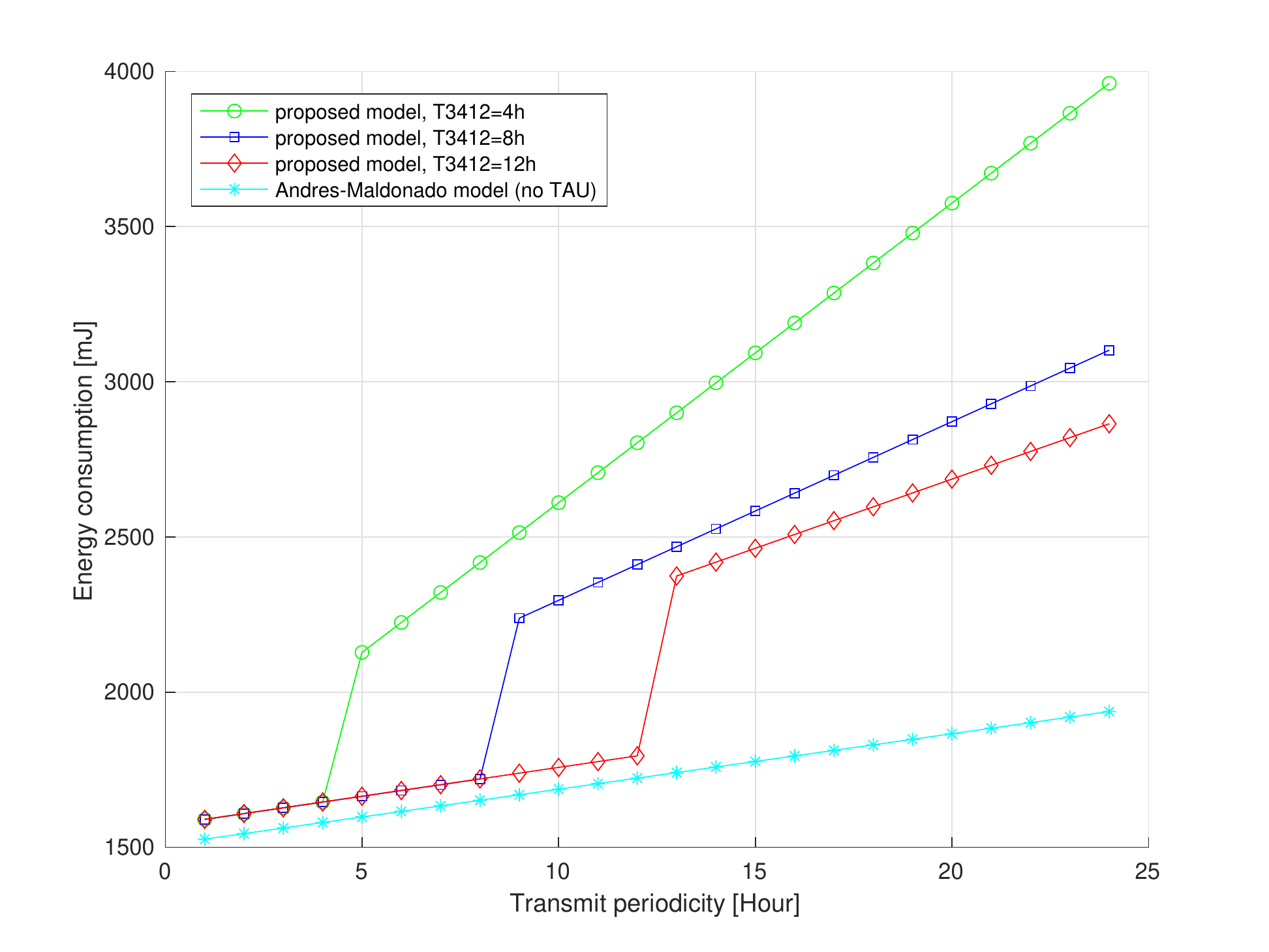}
    \caption{Comparisons of the energy consumption model proposed in \cite{anamodnbiot} with different transmit cycles and $T_{\text{3412}}$ timer, assuming in scenario good and a payload size of 100 Bytes}
    \label{fig:tau_impact}
\end{figure}

A comparison between the proposed energy consumption model and the model proposed in \cite{anamodnbiot} is depicted in \autoref{fig:tau_impact} with different transmit cycles and $T_{\text{3412}}$ timer, assuming in scenario good and a payload size of 100 Bytes. There are similarities between the two models in modelling the energy consumption of the \cgls{ue} states and behaviors, e.g., transmitting/receiving packets or signaling. The difference is that the calculation of the total energy consumption in our proposed model is based on the composition of the associated \cgls{ue} procedures according to the desired \cgls{ue} behavior, while the calculation in \cite{anamodnbiot} relies on the Markov chain analysis. The other main difference between the two models is that the \cgls{tau} procedure is included in our model while it is not considered in \cite{anamodnbiot}. It can be seen from \autoref{fig:tau_impact} that the two models matches quite well when there is no \cgls{tau}. However, the difference is much obvious when the \cgls{tau} procedure occurred. From the measurement it is observed that the occurrence of the \cgls{tau} procedure can be energy expensive and therefore should be carefully configured (i.e., $T_{\text{3412}}$ timer) by the network operator.

\autoref{fig:batterylife_compare} shows the estimated battery lifetime for both the \cgls{nbiot} device N211 and the \cgls{ltem} device R410M under different transmit cycles and  coverage scenarios, assuming the payload size to be 100 Bytes and $T_{\text{3412}}$ timer to be 4 hours. The curve for the \cgls{ltem} device R410M in the extreme scenario is missing as \cgls{ltem} can not reach the area with 160 dB coupling loss. The estimated battery lifetime of N211 is longer than R410M, due to the lower energy or power cost of N211 in each state as compared to R410M. When the transmit cycle is $\geq 24$ hours, the battery lifetime for NB-IoT device N211 can reach up to 10 years, satisfying the 10-year battery lifetime requirement specified by 3GPP.

\subsection{Discussions and Future Work}
\begin{figure}[t]
    \centering
    \includegraphics[width=\linewidth]{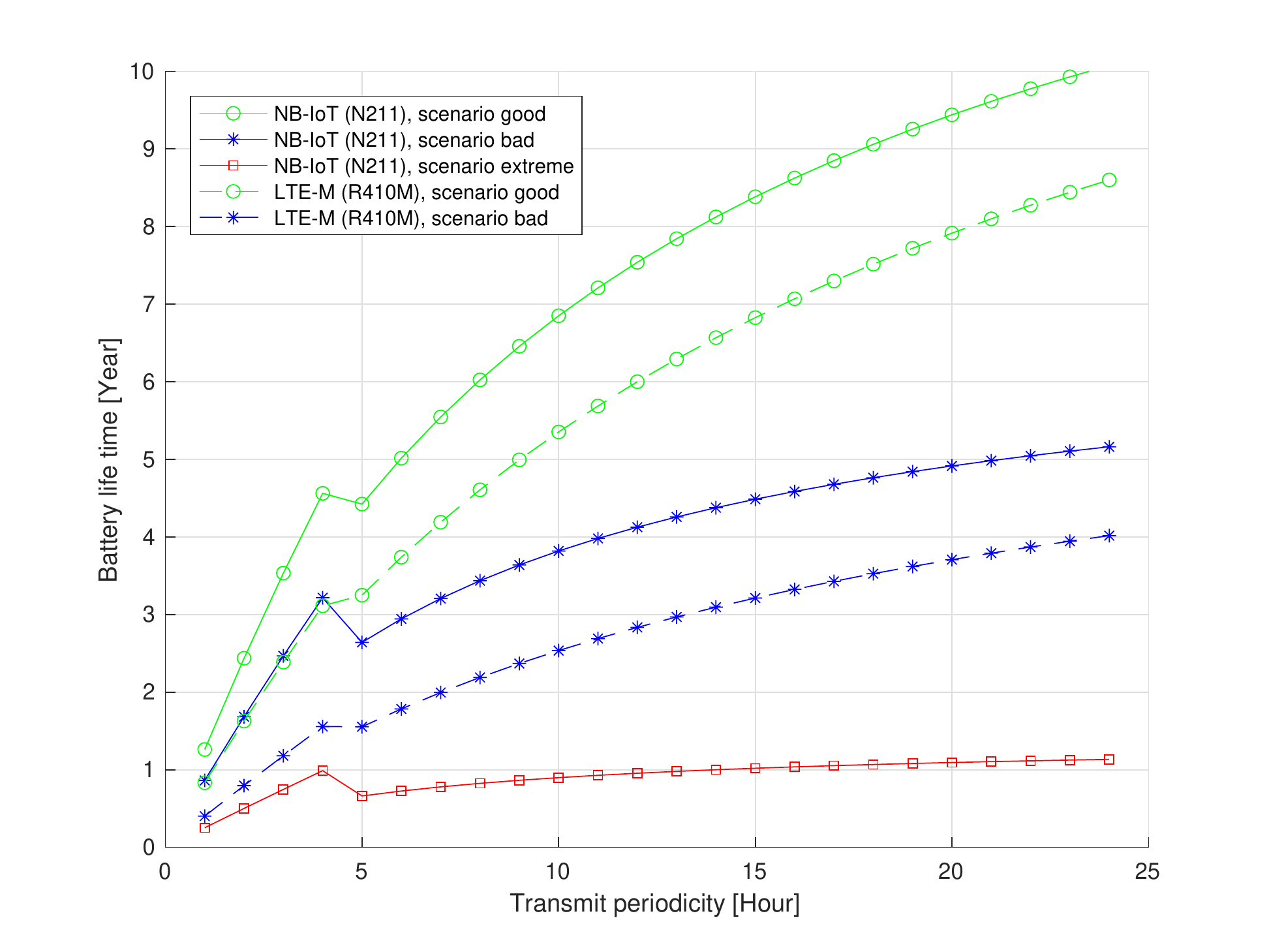}
    \caption{Estimated battery lifetime for both the NB-IoT device N211 and the LTE-M device R410M under different transmit cycles and  coverage scenarios, assuming the payload size to be 100 Bytes and $T_{\text{3412}}$ timer to be 4 hours}
    \label{fig:batterylife_compare}
\end{figure}

It should be noted that the battery lifetime comparison shown in \autoref{fig:batterylife_compare} is only based on the two specific \cgls{nbiot} and \cgls{ltem} devices. The comparison between those two devices can not be generalized to a conclusion that the \cgls{nbiot} device can last for longer time than the \cgls{ltem} device, as the energy footprint of a device is implementation specific. However, the proposed energy consumption model and the battery lifetime estimation method can be generalized to all devices operating with either \cgls{nbiot} or \cgls{ltem}. In order to apply the proposed model for a new \cgls{nbiot} or \cgls{ltem} device, one has to measure and characterize the power consumption in each state of that specific device, as listed in \autoref{tab:PowerEnergyConsumption}. Typically this modem characterization information is provided by the vendors.

From \autoref{fig:lifetime_payload} and \autoref{fig:lifetime_cycle}, it is shown that the battery lifetime of a device not only depends on the power consumption of the device in each state, but also depends on the traffic profile, the coverage scenario, as well as the network configuration parameters. While an \cgls{iot} application developer can adjust the frequency and size of the payload data to be sent towards the server, the coverage and network configuration parameters including power saving parameters are normally set by the network operators. It is found during the study that some of the network configuration parameters, such as the $T_{\text{3412}}$ timer which determines the \cgls{tau} periodicity, have a great impact on the energy efficiency. Increasing the $T_{\text{3412}}$ timer can decrease the energy consumption due to fewer occurrence of the \cgls{tau} procedure and the increased sleep time, at the expense of longer response time. A minimum setting of 240 minutes for the $T_{\text{3412}}$ timer is recommended by GSMA \cite{gsma_ltem}. As a general guideline, it is of crucial importance not only for the \cgls{iot} application developer to carefully select the traffic parameters that best tradeoff between energy efficiency and performance metric, but also for the network operators to fine tune the coverage and the network parameters to ensure a long battery lifetime. With proper configuration of these parameters, the battery lifetime of an \cgls{iot} device can last for 10 years as required by 3GPP.

This paper only considers the energy consumption model for \cgls{nbiot} and \cgls{ltem}. Other \cgls{lpwan} technologies such as Sigfox and LoRa could also be interesting to model and compare with \cgls{nbiot} and \cgls{ltem}. In addition, only the power consumption model for the modem is considered in this paper. Other hardware such as the sensors, the actuators and the processor also need to be taken into account when estimating the battery lifetime. For the traffic model, only deterministic traffic with periodic transmission is assumed. Other traffic models such as non-deterministic traffic are also of interest. Furthermore, for accurate estimation of the battery lifetime, the capacity  leakage of the battery should also be taken into account instead of assuming an ideal battery without capacity leakage. Those could be the future work for modelling the energy consumption of \cgls{lpwan} \cgls{iot} devices.

%% file: 05_Conclusion.tex
\section{Conclusion}\label{sec:conclusion}
This paper presented a comprehensive energy consumption model for \cgls{iot} device battery lifetime estimation, focusing on 3GPP standardized \cgls{lpwan} technologies \cgls{nbiot} and \cgls{ltem}. We start with the introduction of the \cgls{ue} states and procedures, followed by the detailed energy consumption modelling for each \cgls{ue} state and the main procedures. By composing the associated \cgls{ue} states and procedures, the energy consumption of any \cgls{ue} behaviour can be calculated. Besides, a traffic profile which resembles the uplink traffic of most \cgls{iot} applications and three coverage scenarios which determine the physical layer transmission parameters with respect to different coverage levels have also been presented. Once the modem energy consumption model, the traffic profile, and the coverage scenario have been determined, the energy consumption of an \cgls{iot} device within a transmit cycle can be calculated and the corresponding battery lifetime can be estimated. A measurement testbed has been set up to validate the proposed energy consumption model with two commercial \cgls{nbiot} and \cgls{ltem} devices, namely U-Blox EVK-N211 and U-Blox EVK-R410M. The results show that the proposed energy consumption model matches very well with the measurement results in different configurations, with the estimation error within 5\%. The impact of the traffic profile, the coverage scenario, as well as the network configuration parameters on the device's battery lifetime has also been analyzed, showing that both the application specific and the network specific parameters are of crucial importance to ensure a long battery lifetime.

%% file: 00_main.bbl
\begin{thebibliography}{10}
\providecommand{\url}[1]{#1}
\csname url@samestyle\endcsname
\providecommand{\newblock}{\relax}
\providecommand{\bibinfo}[2]{#2}
\providecommand{\BIBentrySTDinterwordspacing}{\spaceskip=0pt\relax}
\providecommand{\BIBentryALTinterwordstretchfactor}{4}
\providecommand{\BIBentryALTinterwordspacing}{\spaceskip=\fontdimen2\font plus
\BIBentryALTinterwordstretchfactor\fontdimen3\font minus
  \fontdimen4\font\relax}
\providecommand{\BIBforeignlanguage}[2]{{%
\expandafter\ifx\csname l@#1\endcsname\relax
\typeout{** WARNING: IEEEtran.bst: No hyphenation pattern has been}%
\typeout{** loaded for the language `#1'. Using the pattern for}%
\typeout{** the default language instead.}%
\else
\language=\csname l@#1\endcsname
\fi
#2}}
\providecommand{\BIBdecl}{\relax}
\BIBdecl

\bibitem{ericssonMobilityReport}
{Ericsson}, ``{Ericsson Mobility Report},''
  \url{https://www.ericsson.com/4adc87/assets/local/mobility-report/documents/2020/november-2020-ericsson-mobility-report.pdf},
  {2020}.

\bibitem{TR45820}
3GPP, ``{Cellular System Support for Ultra Low Complexity and Low Throughput
  Internet of Things; (Release 13)},'' {3rd Generation Partnership Project
  (3GPP)}, {Technical Report (TR)} 45.820, {Version 2.0.0}.

\bibitem{iotbook}
H.~{Wang}, A.~{Sørensen}, M.~{Remy}, N.~{Kjettrup}, J.~J. {Nielsen}, and G.~M.
  {Madue\~{n}o}, \emph{Wireless Networks and Industrial IoT}.\hskip 1em plus
  0.5em minus 0.4em\relax Springer, 2021, ch. Power Measurement Framework for
  LPWAN IoT, pp. 105--129.

\bibitem{mads}
M.~{Lauridsen}, ``\BIBforeignlanguage{English}{{Studies on Mobile Terminal
  Energy Consumption for LTE and Future 5G}},''
  \emph{\BIBforeignlanguage{English}{PhD thesis, Aalborg University}}, Jan.
  2015.

\bibitem{Soussi}
M.~{El Soussi}, P.~{Zand}, F.~{Pasveer}, and G.~{Dolmans}, ``Evaluating the
  performance of emtc and nb-iot for smart city applications,'' in \emph{IEEE
  International Conference on Communications (ICC)}, May 2018, pp. 1--7.

\bibitem{ltemnbiotcovcap}
M.~{Lauridsen}, I.~Z. {Kovacs}, P.~{Mogensen}, M.~{Sorensen}, and S.~{Holst},
  ``Coverage and capacity analysis of lte-m and nb-iot in a rural area,'' in
  \emph{2016 IEEE 84th Vehicular Technology Conference (VTC-Fall)}, Sep. 2016,
  pp. 1--5.

\bibitem{Vejlgaard}
B.~{Vejlgaard}, M.~{Lauridsen}, H.~{Nguyen}, I.~Z. {Kovacs}, P.~{Mogensen}, and
  M.~{Sorensen}, ``Coverage and capacity analysis of sigfox, lora, gprs, and
  nb-iot,'' in \emph{IEEE 85th Vehicular Technology Conference (VTC Spring)},
  June 2017, pp. 1--5.

\bibitem{GermanPaper}
M.~{Lauridsen}, R.~{Krigslund}, M.~{Rohr}, and G.~{Madueno}, ``An empirical
  nb-iot power consumption model for battery lifetime estimation,'' in
  \emph{2018 IEEE 87th Vehicular Technology Conference (VTC-Spring)}, June
  2018, pp. 1--5.

\bibitem{nbiotpsmedrx}
A.~K. {Sultania}, P.~{Zand}, C.~{Blondia}, and J.~{Famaey}, ``Energy modeling
  and evaluation of nb-iot with psm and edrx,'' in \emph{2018 IEEE Globecom
  Workshops (GC Wkshps)}, Dec 2018, pp. 1--7.

\bibitem{Azari}
A.~{Azari}, C.~{Stefanović}, P.~{Popovski}, and C.~{Cavdar}, ``On the
  latency-energy performance of nb-iot systems in providing wide-area iot
  connectivity,'' \emph{IEEE Transactions on Green Communications and
  Networking}, vol.~4, no.~1, pp. 57--68, Oct. 2020.

\bibitem{anamodnbiot}
P.~{Andres-Maldonado}, M.~{Lauridsen}, P.~{Ameigeiras}, and J.~M.
  {Lopez-Soler}, ``Analytical modeling and experimental validation of nb-iot
  device energy consumption,'' \emph{IEEE Internet of Things Journal}, vol.~6,
  no.~3, pp. 5691--5701, June 2019.

\bibitem{nbiot}
P.~{Andres-Maldonado}, P.~{Ameigeiras}, J.~{Prados-Garzon}, J.~{Navarro-Ortiz},
  and J.~M. {Lopez-Soler}, ``Narrowband iot data transmission procedures for
  massive machine-type communications,'' \emph{IEEE Network}, vol.~31, no.~6,
  pp. 8--15, November 2017.

\bibitem{Hertlein}
M.~{Hertlein}, S.~{Breun}, G.~{Cappel}, A.~{Schwarzmeier}, F.~{Lurz},
  R.~{Weigel}, and G.~{Fischer}, ``Evaluation of cellular standards for low
  data rate applications regarding power consumption and timing parameters,''
  in \emph{IEEE Radio and Wireless Symposium (RWS)}, Jan. 2018, pp. 1--3.

\bibitem{Duhovnikov}
S.~{Duhovnikov}, A.~{Baltaci}, D.~{Gera}, and D.~A. {Schupke}, ``Power
  consumption analysis of nb-iot technology for low-power aircraft
  applications,'' in \emph{IEEE 5th World Forum Internet Things (WF-IoT)},
  April 2019, pp. 719--723.

\bibitem{Yeoh}
C.~Y. {Yeoh}, A.~B. {Man}, Q.~M. {Ashraf}, and A.~K. {Samingan}, ``Experimental
  assessment of battery lifetime for commercial off-the-shelf nb-iot module,''
  in \emph{IEEE International Conference on Advanced Communication Technology
  (ICACT)}, Feb. 2018, pp. 223--228.

\bibitem{Martinez}
B.~{Martinez}, F.~{Adelantado}, A.~{Bartoli}, and X.~{Vilajosana}, ``Exploring
  the performance boundaries of nb-iot,'' \emph{IEEE Internet of Things
  Journal}, vol.~6, no.~3, pp. 5702--5712, June 2019.

\bibitem{Abbas}
M.~T. {Abbas}, J.~{Eklund}, K.~J. {Grinnemo}, A.~{Brunstrom}, S.~{Alfredsson},
  O.~{Alay}, S.~{Katona}, G.~{Seres}, and B.~{Rathonyi}, ``Guidelines for an
  energy efficient tuning of the nb-iot stack,'' in \emph{IEEE 45th LCN
  Symposium on Emerging Topics in Networking (LCN Symposium)}, Nov. 2020, pp.
  1--10.

\bibitem{Michelinakis}
F.~{Michelinakis}, A.~S. {Al-Selwi}, M.~{Capuzzo}, A.~{Zanella}, K.~{Mahmood},
  and A.~{Elmokashfi}, ``Dissecting energy consumption of nb-iot devices
  empirically,'' \emph{IEEE Internet of Things Journal}, vol.~8, no.~2, pp.
  1224--1242, Aug. 2020.

\bibitem{Lin}
Y.~B. {Lin}, H.~C. {Tseng}, Y.~W. {Lin}, and L.~J. {Chen}, ``Nb-iottalk: A
  service platform for fast development of nb-iot applications,'' \emph{IEEE
  Internet of Things Journal}, vol.~6, no.~1, pp. 928--939, Feb. 2019.

\bibitem{Ghosh}
N.~{Mangalvedhe}, R.~{Ratasuk}, and A.~{Ghosh}, ``Nb-iot deployment study for
  low power wide area cellular iot,'' in \emph{IEEE 27th Annual International
  Symposium on Personal, Indoor, and Mobile Radio Communications (PIMRC)}, Sep.
  2016, pp. 1--5.

\bibitem{coverageAnalysisLTEM}
{G. Vos, J. Bergman, Y. Bitran, M. Beale, M. Cannon, R. Holden, Y.S. Chan, R.
  Toledano, R. Bras, T. Wakayama, R. Ratasuk, N. Okubo, K. Park, Y. Akimoto,
  T.H. Siregar and S. Lee}, ``{Coverage Analysis for LTE-M CAT-M1 Devices },''
  \emph{Sierra Wireless White Paper}, 2017, (Accessed on 03/18/2019).

\bibitem{Kanj}
M.~{Kanj}, V.~{Savaux}, and M.~L. {Guen}, ``A tutorial on nb-iot physical layer
  design,'' \emph{IEEE Communications Surveys \& Tutorials}, vol.~22, no.~4,
  pp. 2408--2446, Sep. 2020.

\bibitem{TS36213}
3GPP, ``{LTE; Evolved Universal Terrestrial Radio Access (E-UTRA); Physical
  layer procedures},'' {3rd Generation Partnership Project (3GPP)}, {Technical
  Specification (TS)} 36.213, {Version 13.8.0}.

\bibitem{TS36211}
3GPP, ``{LTE; Evolved Universal Terrestrial Radio Access (E-UTRA); Physical
  channels and modulation},'' {3rd Generation Partnership Project (3GPP)},
  {Technical Specification (TS)} 36.211, {Version 13.9.0}.

\bibitem{TS36331}
3GPP, ``{LTE; Evolved Universal Terrestrial Radio Access (E-UTRA); Radio
  Resource Control (RRC); Protocol specification},'' {3rd Generation
  Partnership Project (3GPP)}, {Technical Specification (TS)} 36.331, {Version
  13.8.1}.

\bibitem{lte}
{H. Holma and A. Toskala}, \emph{{LTE for UMTS: Evolution to
  LTE-Advanced}}.\hskip 1em plus 0.5em minus 0.4em\relax {Wiley}, {2012}, no.
  {ISBN: 978-0-470-66000-3}.

\bibitem{LTESim}
{A.S. Pagès}, ``{Link Level Performance Evaluation and Link Abstraction for
  LTE/LTE-Advanced Downlink},'' 2015.

\bibitem{Mocnej}
{J. Mocnej, A. Pekar, W. K. Seah, and I. Zolotova}, ``{Network Traffic
  Characteristics of the IoT Application Use Cases},'' \emph{Technical Report
  Series}, 2018.

\bibitem{gsma_ltem}
{GSMA}, ``{LTE-M Deployment Guide to Basic Feature Set Requirements},''
  \url{https://www.gsma.com/iot/wp-content/uploads/2019/08/201906-GSMA-LTE-M-Deployment-Guide-v3.pdf},
  {2019}.

\end{thebibliography}
